\documentclass[superscriptaddress,showkeys,twocolumn,nofootinbib,longbibliography]{revtex4-1}

\newcommand{\beq}{\begin{equation}}
\newcommand{\beqn}{\begin{eqnarray}}
\newcommand{\eeq}{\end{equation}}
\newcommand{\eeqn}{\end{eqnarray}}

\usepackage{amsmath}
\usepackage{amssymb}
\usepackage{graphicx}
\usepackage{epsf}
\usepackage{slashed}
\usepackage{enumitem}
\usepackage[czech,english]{babel}
\usepackage[cp1250]{inputenc}
\usepackage{hyperref}
\usepackage{xcolor}
\usepackage{physics}

\usepackage[only,llbracket,rrbracket,llparenthesis,rrparenthesis]{stmaryrd} 
\usepackage{accsupp} 

\usepackage[font=small,labelfont=bf,justification=raggedright]{caption}
\usepackage{subcaption}

\usepackage{bbm} 

\usepackage[nodayofweek]{datetime}
\newdateformat{mydate}{\twodigit{\THEDAY}{ }\shortmonthname[\THEMONTH], \THEYEAR}
\usepackage[normalem]{ulem}

\usepackage{color}
\usepackage{ulem} 

\definecolor{myred}{rgb}{1,0,0}
\definecolor{mygreen}{rgb}{0,0.8,0.2}
\definecolor{myblue}{rgb}{0,0,1}

\definecolor{Ared}{rgb}{1,0.7,0}
\definecolor{Agreen}{rgb}{0.7,0.8,0.2}
\definecolor{Ablue}{rgb}{0,0.7,1}

\renewcommand{\emph}[1]{\textit{#1}}

\begin{document}


\title{Standard Model Effective Field Theory and Oscillons}

\author{Z. Drogosz}

\affiliation{Institute of Theoretical Physics, Jagiellonian University,
Lojasiewicza 11, Krak\'{o}w, Poland}

\author{E.I. Sfakianakis}

\affiliation{Texas Center for Cosmology and Astroparticle Physics, Weinberg Institute for Theoretical Physics, Department of Physics, The University of Texas at Austin, Austin, TX 78712, USA} \affiliation{Department of Physics, Harvard University, Cambridge, MA, 02131, USA}

\author{K. Slawinska}

\affiliation{Institute of Theoretical Physics, Jagiellonian University,
Lojasiewicza 11, Krak\'{o}w, Poland}

\author{A. Wereszczynski}

\affiliation{Institute of Theoretical Physics, Jagiellonian University,
Lojasiewicza 11, Krak\'{o}w, Poland}
\affiliation{International Institute for Sustainability with Knotted Chiral Meta Matter (WPI-SKCM$^{\; 2}$), Hiroshima University, 1-3-1 Kagamiyama, Higashi-Hiroshima, Hiroshima 739-8526, JAPAN}

\begin{abstract}
We show that the inclusion of a dimension-six operator in the Higgs potential has a dramatic impact on the stability of oscillons in the $SU(2)$ bosonic sector of the Standard Model, extending their lifetime by orders of magnitude. This happens for the physical value of the ratio between the Higgs and W boson masses, $m_H/m_W=1.556$ 
and for the dimension-six operator $\mathcal{O}_6 = (\Phi^\dagger \Phi)^3$ whose coupling constant is below the current upper bound. 
\end{abstract}

\maketitle

\section{Motivation}

Although there are no stable topological solitons in electroweak theory (EW), both the nontrivial topological structure of the configuration space and the nonlinearities of the field equations give rise to nonperturbative soliton-like solutions. The first is the famous {\it sphaleron} \cite{Manton:1983nd, Klinkhamer:1984di} which, introducing baryon-number-violating processes, plays an important role in baryogenesis. 

The second, much less explored class of solutions consists of oscillons~\cite{Bogolyubsky:1976nx, Gleiser:1993pt}: long-lived, spatially localized, oscillating field configurations that can arise in a wide variety of nonlinear field theories~\cite{Copeland:1995fq, Arodz:2007jh, Gleiser:2011xj, Amin:2011hj, Lozanov:2014zfa, Zhang:2020bec, Olle:2020qqy, Zhou:2024mea,Navarro-Obregon:2023hqe, Blaschke:2024dlt}. 
Unlike topological solitons, oscillons are not protected by topology. Instead, their longevity is a dynamical consequence of nonlinear interactions, which allow the configuration to remain localized.
It has recently been shown that an oscillon is a localized non-normalizable resonance (a threshold/anti-bound mode), where localization is due to the nonlinearity of the model \cite{Blaschke:2026vxj}. Oscillons are especially relevant for the evolution of the early Universe, where, e.g.~they are expected to be produced after inflation, in phase transitions, or in collisions of cosmic defects \cite{Hindmarsh:2006ur, Gleiser:2006te, Saffin:2006yk,Amin:2010jq, Amin:2011hj, Lozanov:2014zfa, Antusch:2015ziz,  Lozanov:2017hjm,  Olle:2019kbo, Amin:2019ums, Iarygina:2020dwe, Hiramatsu:2020obh, Sang:2020kpd,  Kawasaki:2020jnw, Mahbub:2023faw, Aurrekoetxea:2023jwd, Shafi:2024jig, Jia:2024fmo,  Waeming:2026wrp}. 

An oscillon has been constructed in the $SU(2)$ \cite{Farhi:2005rz, Sfakianakis:2012bq} and $SU(2)\times U(1)$ \cite{Graham:2006vy} sectors of the EW theory. However, its long lifetime quickly diminishes once the ratio of Higgs and W masses moves away from the ratio $m_H/m_W=2$.  Since the value realized in nature is $m_H/m_W\simeq 1.556$, this  coherent multi-particle excitation remains mathematically interesting albeit not physically realized.

Despite the success of the   Standard Model (SM) and its remarkable agreement with experimental data, there are still some open questions that are not addressed solely within the SM, e.g.~electroweak first order phase transition and baryogenesis. 
This motivates the extension of the SM. The simplest phenomenological-like approach is known as Standard Model Effective Field Theories (SMEFT), where one includes higher-dimension non-renormalizable operators \cite{Brivio:2017vri, Isidori:2023pyp, Falkowski:2023hsg}. These are thought to arise from a new fundamental theory and can be in principle computed by integrating the new additional (higher energy) degrees of freedom (particles). In the low energy limit, the SMEFT must  reduce to the well-tested SM. 

The aim of the present work is to investigate the oscillon in the $SU(2)$ sector of the EW theory with the inclusion of a SMEFT-motivated operator. We assume that $g'=0$, which results in the decoupling of the $U(1)$ sector. The impact of non-zero $g'$ on the SM oscillon is expected to be rather small \cite{Farhi:2005rz, Graham:2006vy}. 

In SMEFT, there are 20 different dimension-six operators \cite{Buchmuller:1985jz, Grzadkowski:2010es} that contribute to the solitonic (sphaleron and oscillon) sector \cite{Elias-Miro:2013mua, Gan:2017mcv}. Among them $\mathcal{O}_6 = (\Phi^\dagger \Phi)^3$ is especially interesting, as it can trigger the electroweak first-order phase transition if its coefficient is large enough \cite{Grojean:2004xa, Delaunay:2007wb}. It is the only operator which polynomially depends only on the Higgs field and can thus be considered an extension of the Higgs potential. Here, we will consider only this operator.

\section{Effective $SU(2)$ Higgs model}
We will focus on a truncated sector of the electroweak theory, where the $U(1)$ interaction and fermions are neglected. This is the $SU(2)$ gauged Higgs model 
\begin{equation}
\mathcal L =  \frac{1}{2} \operatorname{tr} (W^{\mu \nu}W_{\mu \nu}) + D_\mu \Phi^\dagger D^\mu \Phi - V(\Phi^\dagger \Phi ), 
\end{equation}
where a Higgs doublet $\Phi$ is coupled to an $SU(2)$ gauge field $W_\mu=(-i/2)\tau^a W^a_\mu$, where $\tau^a$ are Pauli matrices. The field tensor reads $W_{\mu \nu}=\partial_\mu W_\nu - \partial_\nu W_\mu +g[W_\mu,W_\nu]$, while the covariant derivative is $D_\mu \Phi = (\partial_\mu +gW_\mu)\Phi$. 
\begin{figure*}
	\includegraphics[width=0.65\columnwidth]{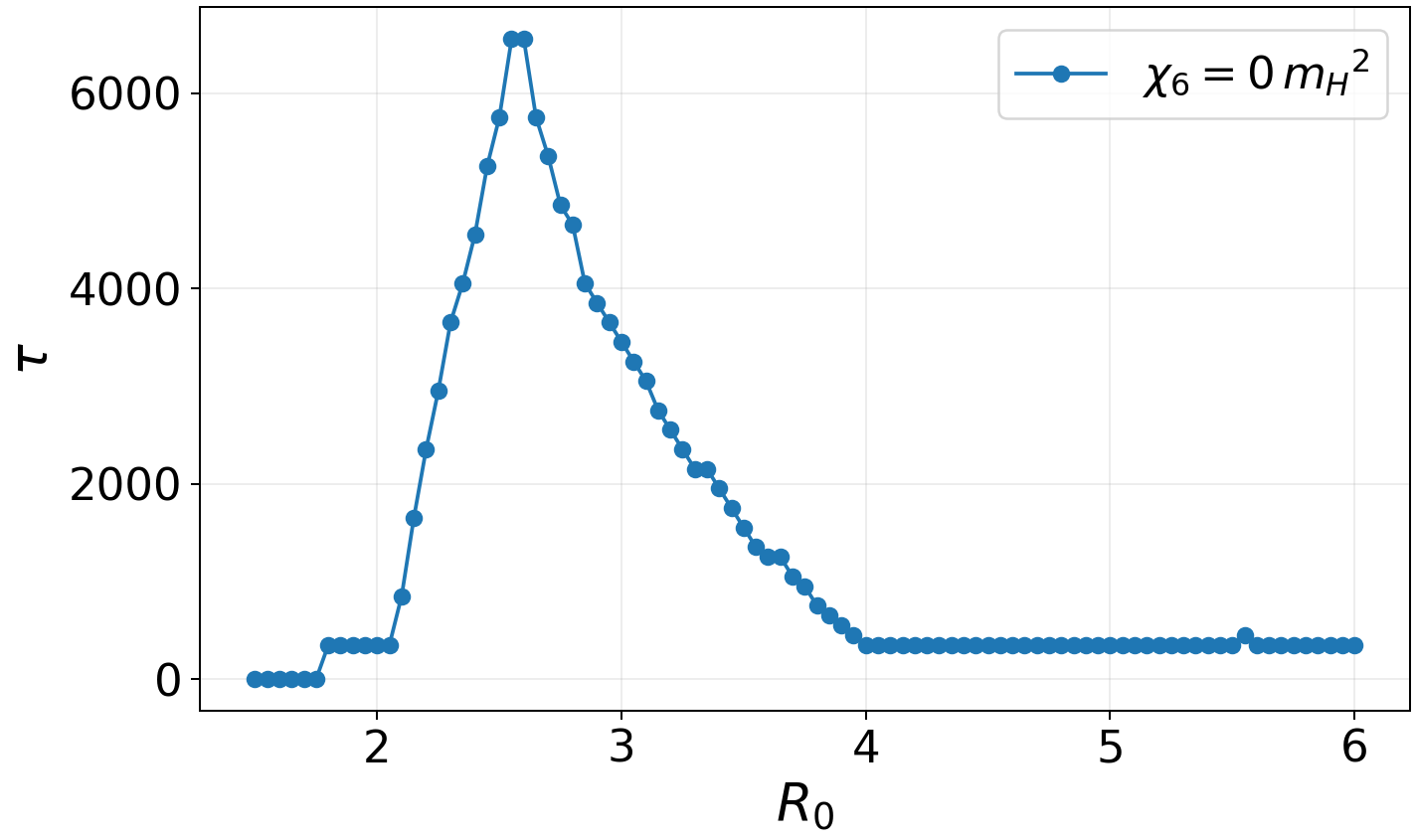} 
    \includegraphics[width=0.65\columnwidth]{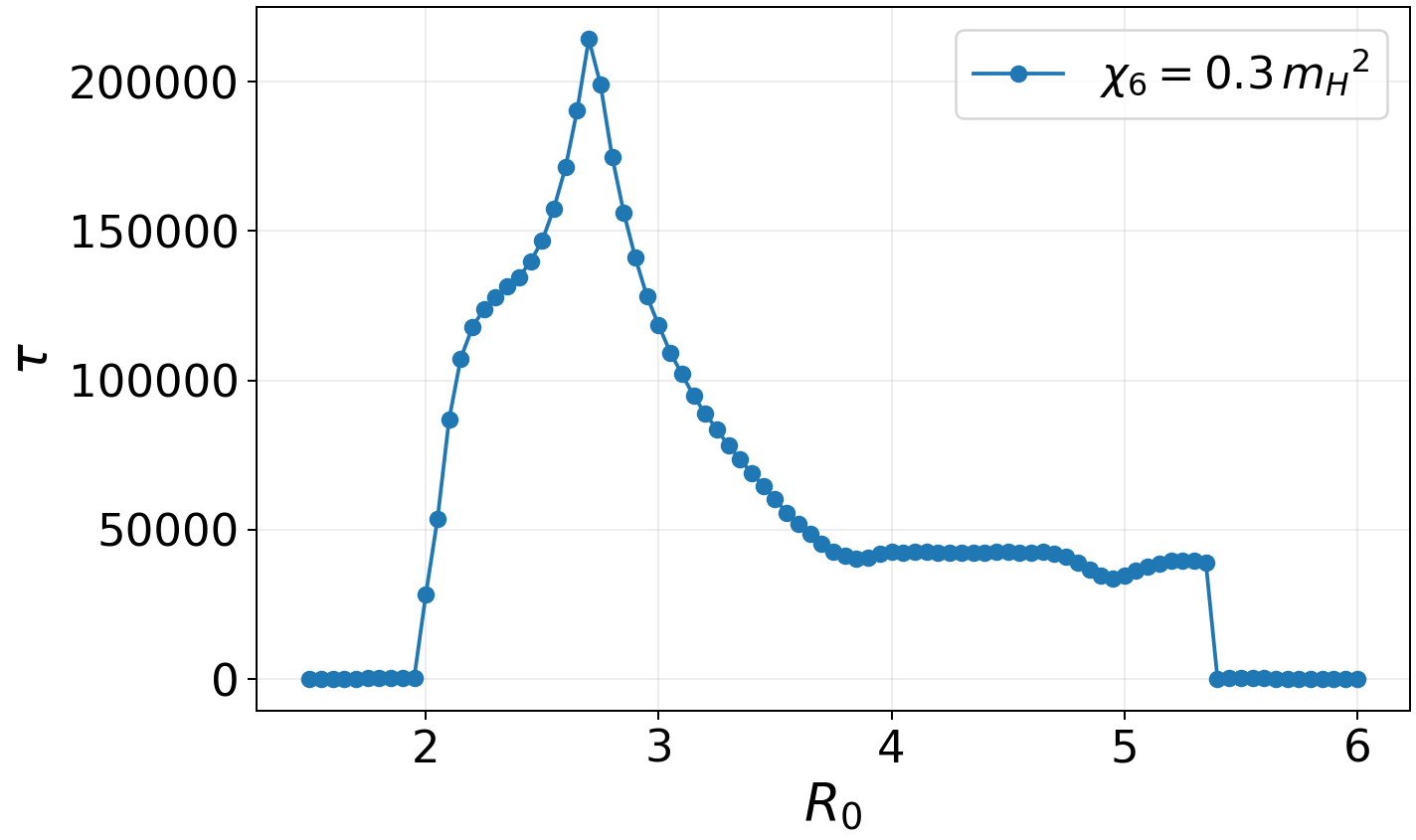}
    \includegraphics[width=0.65\columnwidth]{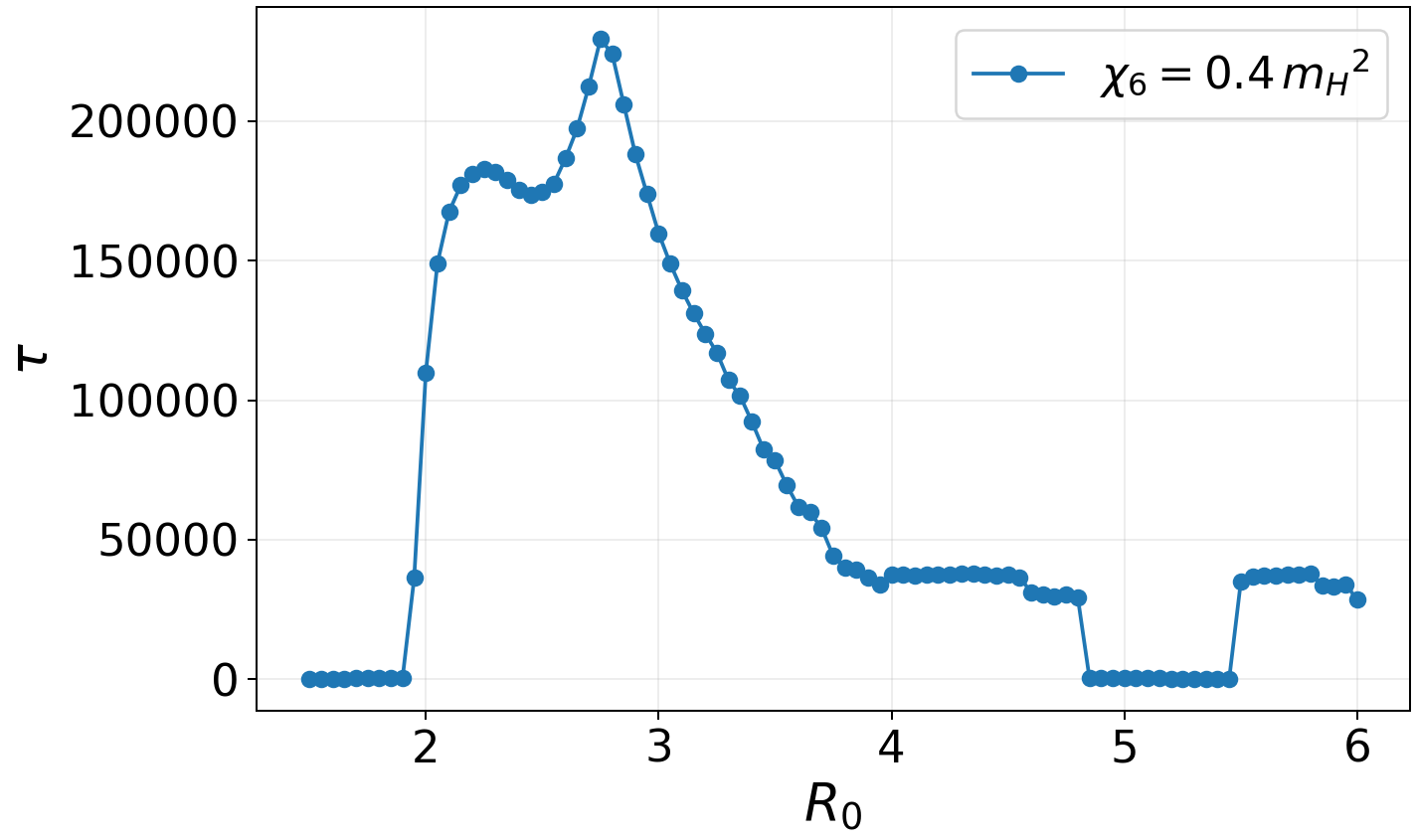}
    \includegraphics[width=0.65\columnwidth]{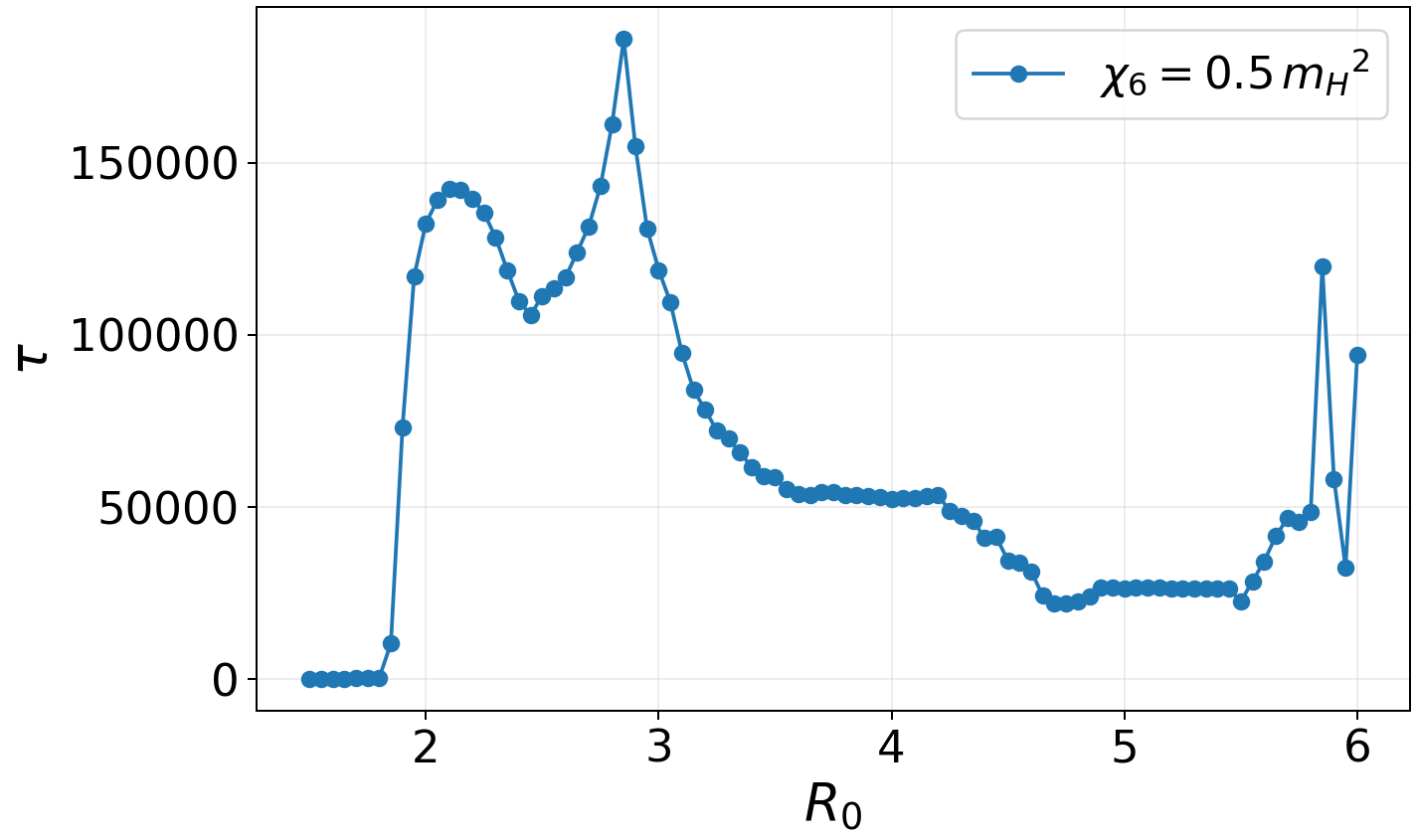}
    \includegraphics[width=0.65\columnwidth]{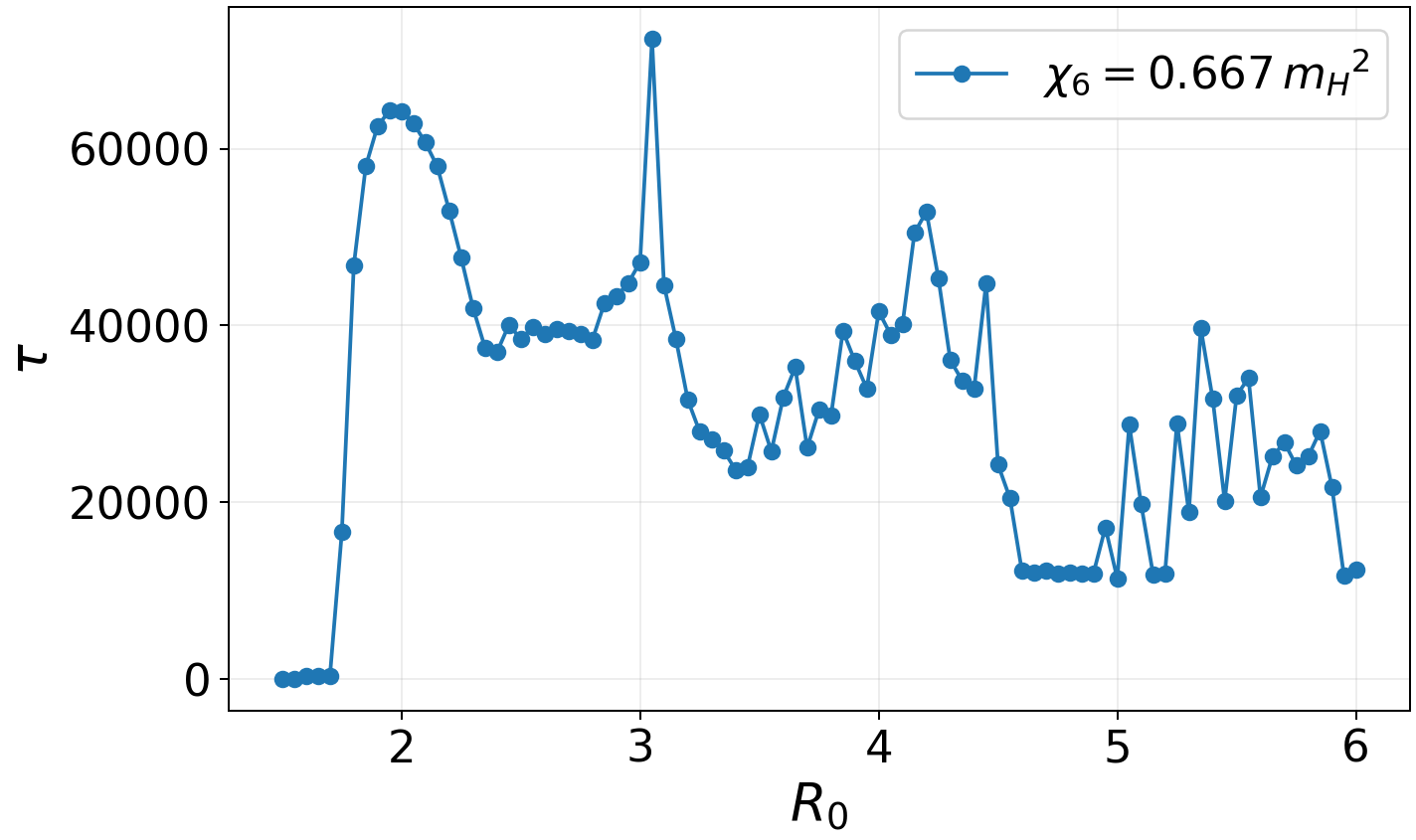}
    \includegraphics[width=0.65\columnwidth]{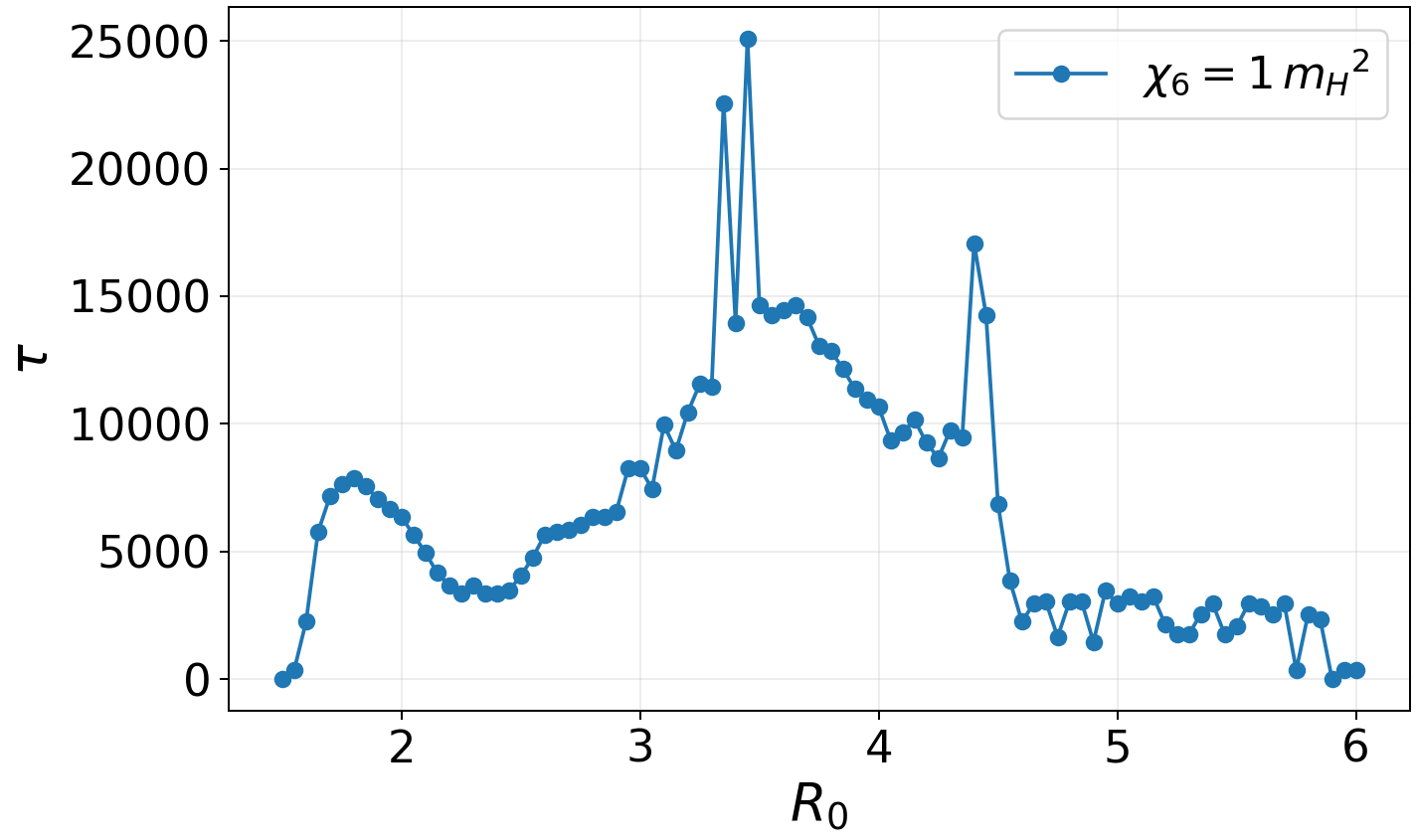}
	\caption{
    Lifetime of the single-field Higgs oscillon versus the initial Gaussian width $R_0$ (see Eq.~\eqref{K-init}) with $A_K=-2$ and $m_H/m_W=1.556$, when the dimension-six operator $(\Phi^\dagger \Phi)^3$ is added to the SM Higgs potential. The $\chi_6=0$ panel corresponds to the quartic theory; nonzero $\chi_6$ dramatically increases the lifetime and broadens the range of $R_0$ leading to longer-lived oscillons.
    } 
    \label{K-oscillon-life}
\end{figure*}

We generalize the usual quartic symmetry-breaking  potential by including the dimension-six operator $(\Phi^\dagger\Phi)^3$ \cite{Gan:2017mcv}. It is convenient to write the potential as
\begin{equation}
V(\Phi)  = \frac{m_H^2}{2v^2}\left(\Phi^\dagger\Phi-\frac{v^2}{2}\right)^2 + \frac{c_6}{\Lambda^2} \left(\Phi^\dagger\Phi-\frac{v^2}{2}\right)^3,
\end{equation}
where $m_H$ is the mass of the Higgs, $v$ defines the broken vacuum, $\Phi^\dagger\Phi=v^2/2$ and the ratio $c_6/\Lambda^2$ parameterizes the dimension-six extension. The mass of the gauge field is still $m_W=vg/2$.

The sixth-order term qualitatively changes the behavior of the potential at $\Phi=0$. This depends on the value of the coupling strength of the new operator, $c_6/\Lambda^2$, which is the ratio of the dimensionless Wilson coefficient $c_6$ and the energy scale $\Lambda$. 
For $c_6/\Lambda^2>(c_6/\Lambda^2)_{\rm crit} = (2/3) m_H^2 /v^4$
a new minimum at $\Phi=0$ appears, which becomes a global minimum at $c_6/\Lambda^2>m_H^2/v^4$.

The equations of motion for the coupled Higgs-W system are
\begin{equation}
     D_\mu D^\mu \Phi + \frac{m^2_H}{v^2} \left( \Phi^\dagger \Phi -\frac{v^2}{2} \right)\Phi+\frac{3c_6}{\Lambda^2} \left( \Phi^\dagger \Phi -\frac{v^2}{2} \right)^2\Phi=0,
\end{equation}
  \begin{equation}
     (D^\mu F_{\mu \nu})^a+\frac{i g^2}{2} \left[\Phi^\dagger \tau^a D_\nu \Phi - (D_\nu\Phi)^\dagger \tau^a \Phi \right]=0.
\end{equation} 
We will study the dynamics of the extended $SU(2)$ Higgs model in the spherical ansatz reduction
\begin{align}
  W_0^a(x) &= \frac{1}{g} G(r,t) \frac{x_a}{r},\\
  W_j^a(x) &= \frac{1}{g}\bigg[
    \frac{f_A(r,t)-1}{r^2} \varepsilon_{jam}\,x_m
    + \nonumber \\
    &\frac{f_B(r,t)}{r^3}\bigl(r^2\delta_{ja}-x_jx_a\bigr)
    + \frac{f_C(r,t)}{r^2} x_jx_a
  \bigg],\\
  \Phi(x) &= \frac{v}{\sqrt{2}}\left[
    H(r,t)
    + i K(r,t) \frac{\boldsymbol{\tau}\cdot\mathbf{x}}{r}
  \right]\begin{pmatrix}0\\1\end{pmatrix},
  \label{eq:Phi_dim6}
\end{align}
where $r=|\mathbf{x}|$.  We work in the temporal gauge, $W_0^a=0$, so $G=0$, and
the dynamical fields are $(f_A,f_B,f_C,H,K)$. In Appendix \ref{app-eom} we show the field equations that result from this ansatz. In our numerics, it is convenient to introduce a new dimensionless parameter
$\chi_6 = 16 c_6 m_W^2/(g^4 \Lambda^2)$. The false vacuum appears for $\chi_6 > \chi_6^{\rm crit}=(2/3) m^2_H/m^2_W$ and becomes degenerate
with the broken minimum at $\chi_6=m^2_H/m^2_W$.

\section{Higgs-channel oscillon}
\label{1-field}

\emph{Single-field sector.---}The simplest oscillon supported by the spherical ansatz  is  obtained by setting all fields to their vacuum values except one Higgs component, $K$. This gives a single real scalar theor with a sextic-self interaction in the Higgs potential, 
which is known to support oscillons in (1+1) and (2+1) dimensions \cite{Fodor:2009kf, Martinez:2026hki}. 
Without loss of generality, we work in units in with $c=1$ and $m_W=1$.
We evolve the e.o.m. for $K(r,t)$ with   Gaussian initial data
\begin{equation}
    K(r,0)=1+A_Ke^{-\frac{r^2}{R_0^2}}, \label{K-init}
\end{equation}
and identify the parameter range of $R_0$ leading to long-lived oscillons.
We choose $A_K=-2$, where the profile has $K(0,0)=-1$ and $K(\infty,0)=1$,
corresponding to a large-amplitude excitation connecting two degenerate
vacuum points in the single-field truncation.

To see the impact of the dimension-six term we plot the lifetime of the generated oscillons as a function of $R_0$ in Fig.~\ref{K-oscillon-life}.  We use the physical mass of Higgs $m_H=1.556$ in units of $m_W=1$ and consider several values of $\chi_6$. 
For $\chi_6=0$ we recover the $\phi^4$ model. The oscillons in $\phi^4$ theory were widely considered in the literature \cite{Honda:2001xg,Fodor:2006zs, Fodor:2008du, Fodor:2009kf, Salmi:2012ta}. The top left panel of Fig. \ref{K-oscillon-life} reproduces the lifetime results of Ref. \cite{Honda:2001xg} to good accuracy, which validates our numerical code.

\begin{figure}
	\includegraphics[width=\columnwidth]{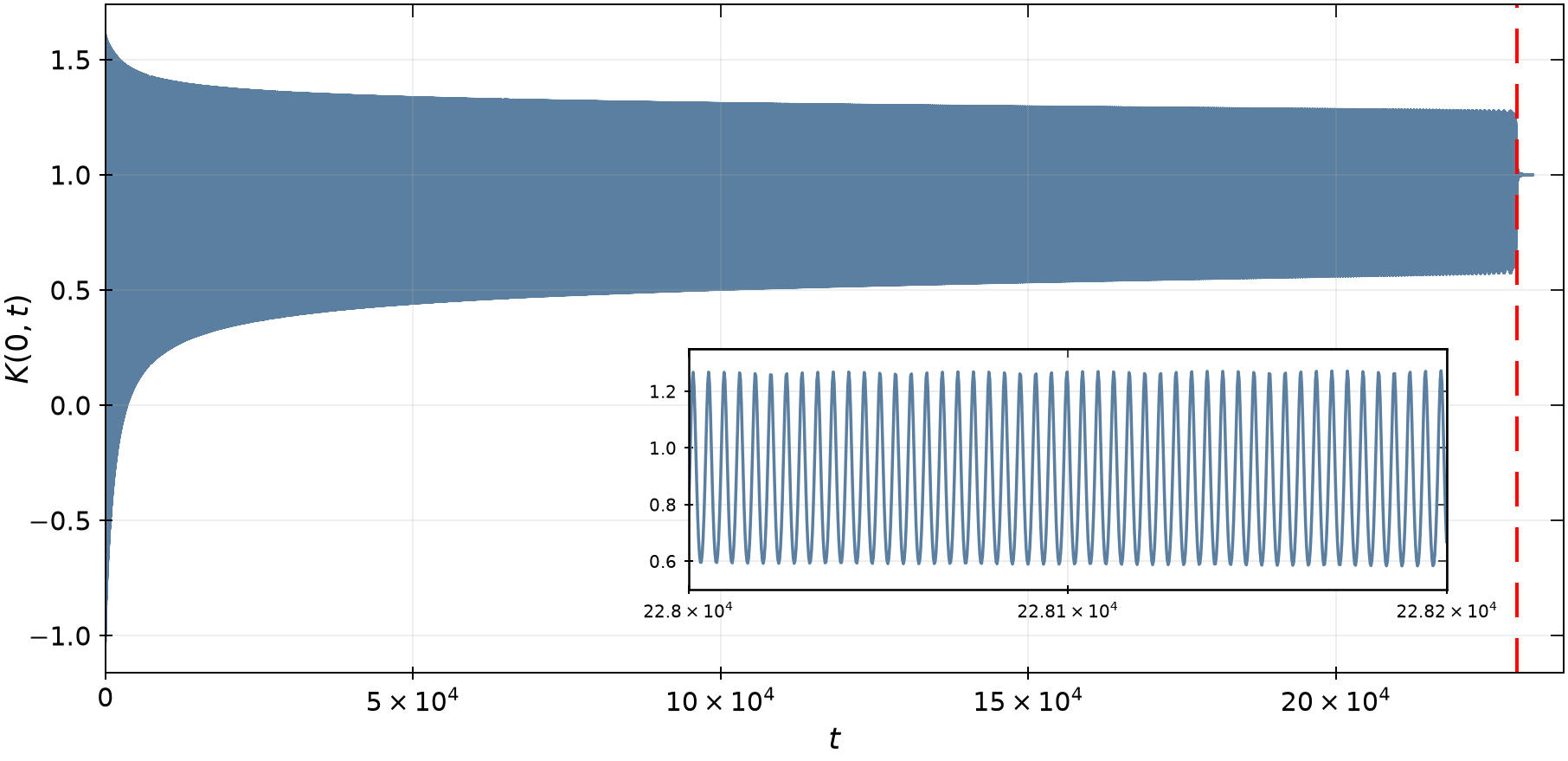} 
	\caption{
    Representative long-lived single-field Higgs oscillon. The plot shows the central Higgs component $K(0,t)$ for initial data $A_K=-2$, $R_0=2.75$, and $\chi_6/m_H^2=0.4$. The lifetime is $\tau\simeq 2.3\times10^5\,m_W^{-1}$, illustrating the enhancement due to the sextic interaction (see Fig.~\ref{K-oscillon-life}). }
    \label{K-oscillon}
\end{figure}

The first important result is that the inclusion of the sextic term can {\it extend the lifetime} by a factor of more than $50$, changing it from $\tau \approx 6\times 10^3$ for $\chi_6=0$ to $\tau > 2 \times 10^5$. 
The lifetime is defined as the time for which the energy stored within a radius that contained the oscillon drops by $90\%$ from a previous plateau level computed as a moving average over a long time window. 
A strong effect is already observed for a relatively small value of $\chi_6 \approx 0.2m_H^2$, where the lifetime is larger than $10^5$ in units of $m_W^{-1}$. An example of the oscillon with lifetime $\tau\approx 230 \, 000$ is in Fig.~\ref{K-oscillon}. Its fundamental frequency increases slowly from $\omega\approx 1.50$ to $\omega\approx 1.52$ in units of $m_W$, just before the oscillon's disintegration.

Furthermore, {\it the basin of attraction for oscillon formation is significantly larger}.
That is, a much larger range of $R_0$ leads to  oscillon formation. In the $\phi^4$ model they appear for $R_0\in (2,4)$, while for $\chi_6=0.2 m_H^2$ we find very long-lived oscillons almost in the whole range $R_0\in [2,6]$.

The second interesting observation is the change in the resonance structure of the lifetime curve. The single peak observed for the $\phi^4$ theory grows and splits into two peaks as $\chi_6$ increases (up to $\chi_6\simeq 0.4$). For larger values of $\chi_6$ increases (up to to $\chi_6\simeq 0.6$), the lifetime decreases slowly while one of the peaks near  $R_0 \approx 3$ becomes very narrow. Around the onset of the additional local potential minimum at $\chi_6=(2/3)m_H^2$, the lifetime curve develops a more intricate multi-peak structure, which simplifies again as the two potential minima approach degeneracy at $\chi_6=m_H^2$.

We must note that the curves in Fig.~\ref{K-oscillon-life} do not resolve the fine comb-like resonance structure known in the $\chi_6=0$ theory from denser scans in
$R_0$~\cite{Honda:2001xg}. These narrow resonances can locally enhance the lifetime, which however remains much shorter than that of the longest-lived oscillons found here with the inclusion of the sextic term. Analogous fine structure is expected
for $\chi_6\neq0$.

\begin{figure}
	\includegraphics[width=\columnwidth]{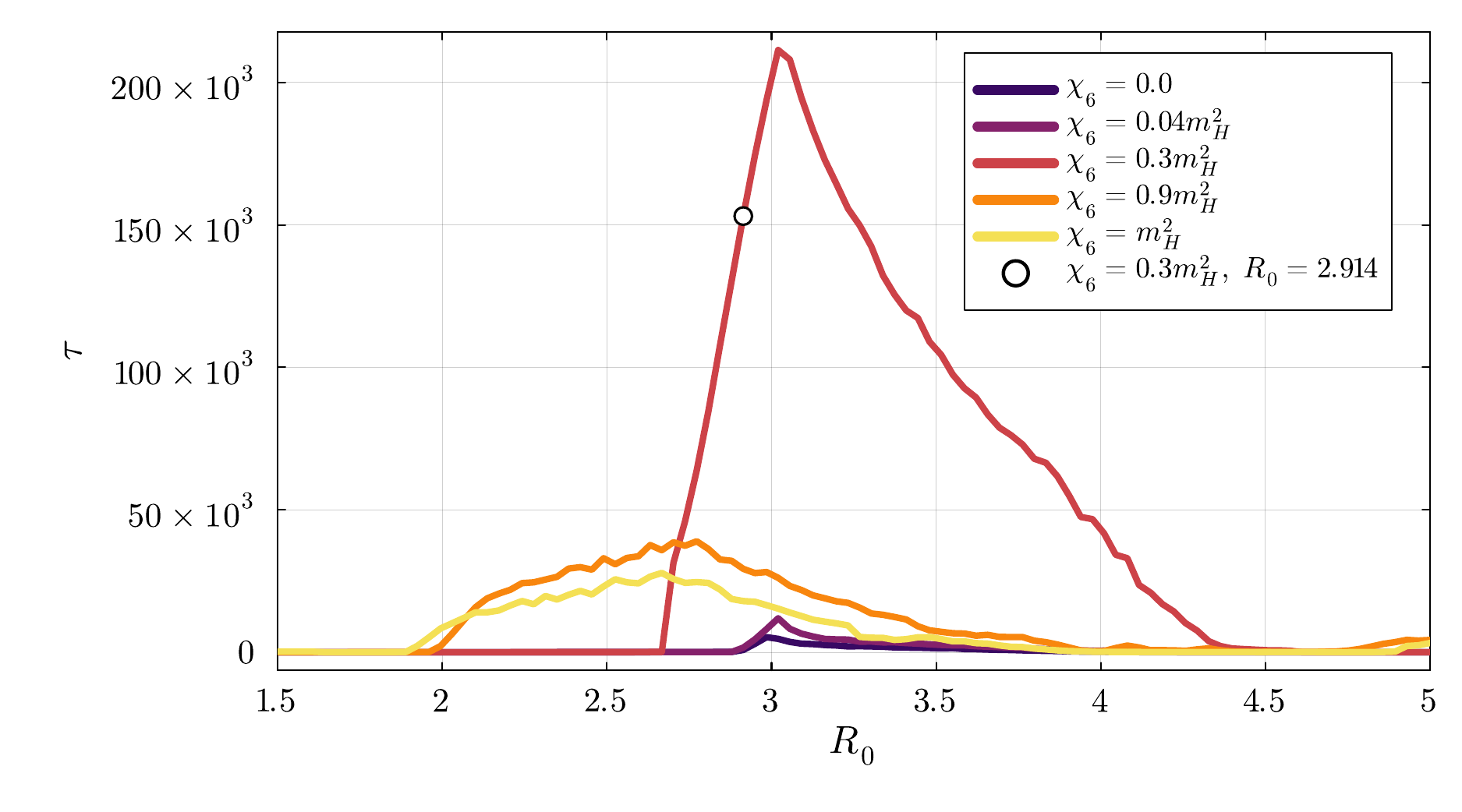} 
	\caption{Oscillon lifetime after opening one gauge-field direction; the gauge
component is initialized following Eq.\eqref{fA-init}.
The long-lived configurations relax to the $K$-dominated Higgs-channel oscillon and the stabilizing effect of the sextic interaction persists away from the strict single-field limit.} 
    \label{2-channel}
    \includegraphics[width=\columnwidth]{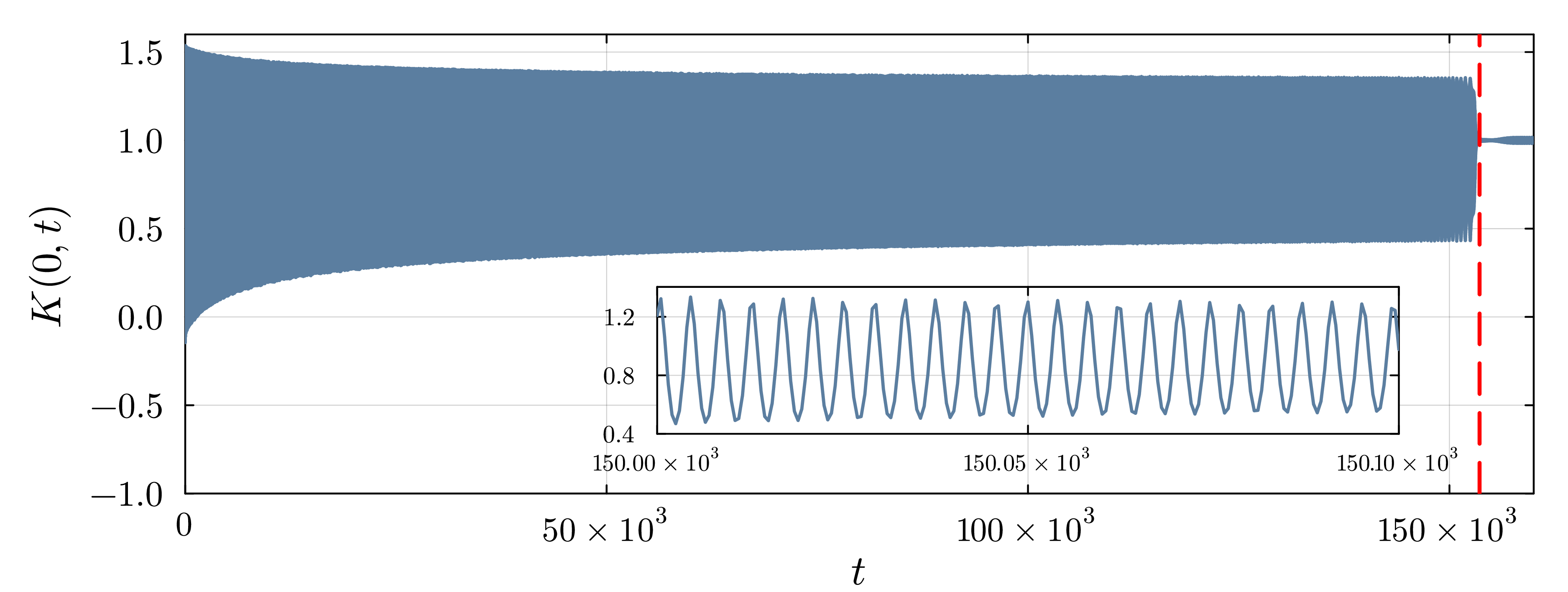} 
	\caption{Representative evolution after opening the $f_A$ gauge-field direction, corresponding to the circled point in Fig.~\ref{2-channel}. The plot
shows $K(0,t)$ for $\chi_6/m_H^2=0.3$, $A_K=-1$, $R_0=2.914$, $A_A=0.1$, and $R_A=2.30$. The gauge perturbation rapidly relaxes toward $f_A=-1$, while the configuration settles into a long-lived $K$-dominated Higgs-channel oscillon with $\tau\simeq1.5\times10^5\,m_W^{-1}$.} 
    \label{2-channel-example}
\end{figure}

\emph{Gauge-field perturbation.---}We next relax the strict single-field truncation by allowing the gauge
component $f_A$ to evolve, while keeping $f_B=f_C=H=0$. This is still a consistent two-field truncation of the $SU(2)$ Higgs theory: If $f_B=f_C=H$ and their time derivatives vanish initially, they  remain zero
under the equations of motion.

This deformation away from the single-field limit could, in principle, destabilize the oscillon. In the single-field sector the oscillon frequency frequency lies below the Higgs mass threshold, $\omega<m_H$, suppressing radiation into
Higgs waves. Once a gauge-field direction is opened, however, the relevant mass threshold is lower, $m_W<m_H$. Since the oscillon frequency is typically above $m_W$, one might expect efficient radiation into the gauge
sector.

We do not observe such a destabilization. This is consistent with similar behavior in the global vortex model, where the angular mode of the complex scalar $\phi=f e^{i\theta}$ is gapless. Even in the presence of this massless channel, long-lived oscillons in the broken vacuum can exist when the potential contains a sufficiently large sextic term, corresponding to the appearance of a false unbroken minimum~\cite{Martinez:2026hki}.

We use the same Gaussian profile for the Higgs component $K$ as in Eq.~\eqref{K-init}, now with $A_K=-1$, and supplement it by a localized perturbation of the gauge component around its vacuum value,
\begin{equation}
    f_A(r,0)=-1+A_A r^2e^{-\frac{r^2}{R_A^2}}\, . \label{fA-init}
\end{equation}
Unless  otherwise stated, we take $A_A=0.1$ and $R_A=2.3$. During the evolution, $f_A$ rapidly relaxes toward its vacuum value, $f_A=-1$. The long-lived configuration is therefore not a genuinely two-field oscillon but rather the Higgs-channel oscillon found above. This indicates that the
single-field Higgs-channel oscillon is not immediately destabilized by moving away from the strict single-field limit.

In Fig.~\ref{2-channel} we show the lifetime obtained after opening the $f_A$ gauge-field direction, as a function of $R_0\in[1.5,5]$. We compare the quartic case, $\chi_6=0$, with $\chi_6/m_H^2=0.04,0.3,0.9, 1$. The first two values lie below the threshold for an additional local minimum at the symmetric point, the third lies above it, and the last corresponds to degenerate symmetric and broken minima.

In Fig.~\ref{2-channel-example} we show a representative oscillon evolution after opening the $f_A$ direction, with lifetime $\tau\simeq 1.5\times 10^5\,m_W^{-1}$. The gauge perturbation decays quickly, while $K(0,t)$ settles into the long-lived oscillatory behavior of the single-field Higgs oscillon described at the start of Sec.~\ref{1-field}.

The results are qualitatively consistent with the single-field case. The lifetime increases substantially as the sextic coupling is turned on; for $\chi_6/m_H^2=0.3$ we find configurations with $\tau>2\times 10^5\,m_W^{-1}$. For larger $\chi_6$, once the symmetric local minimum appears, the lifetime decreases, but it remains significantly longer than in the quartic theory. Thus, opening this gauge-field direction does not remove the stabilizing effect of the dimension-six operator: The system still relaxes to a long-lived, $K$-dominated Higgs-channel oscillon at the physical Higgs-to-$W$ mass ratio, $m_H/m_W=1.556$.

\begin{figure}
	\includegraphics[width=\columnwidth]{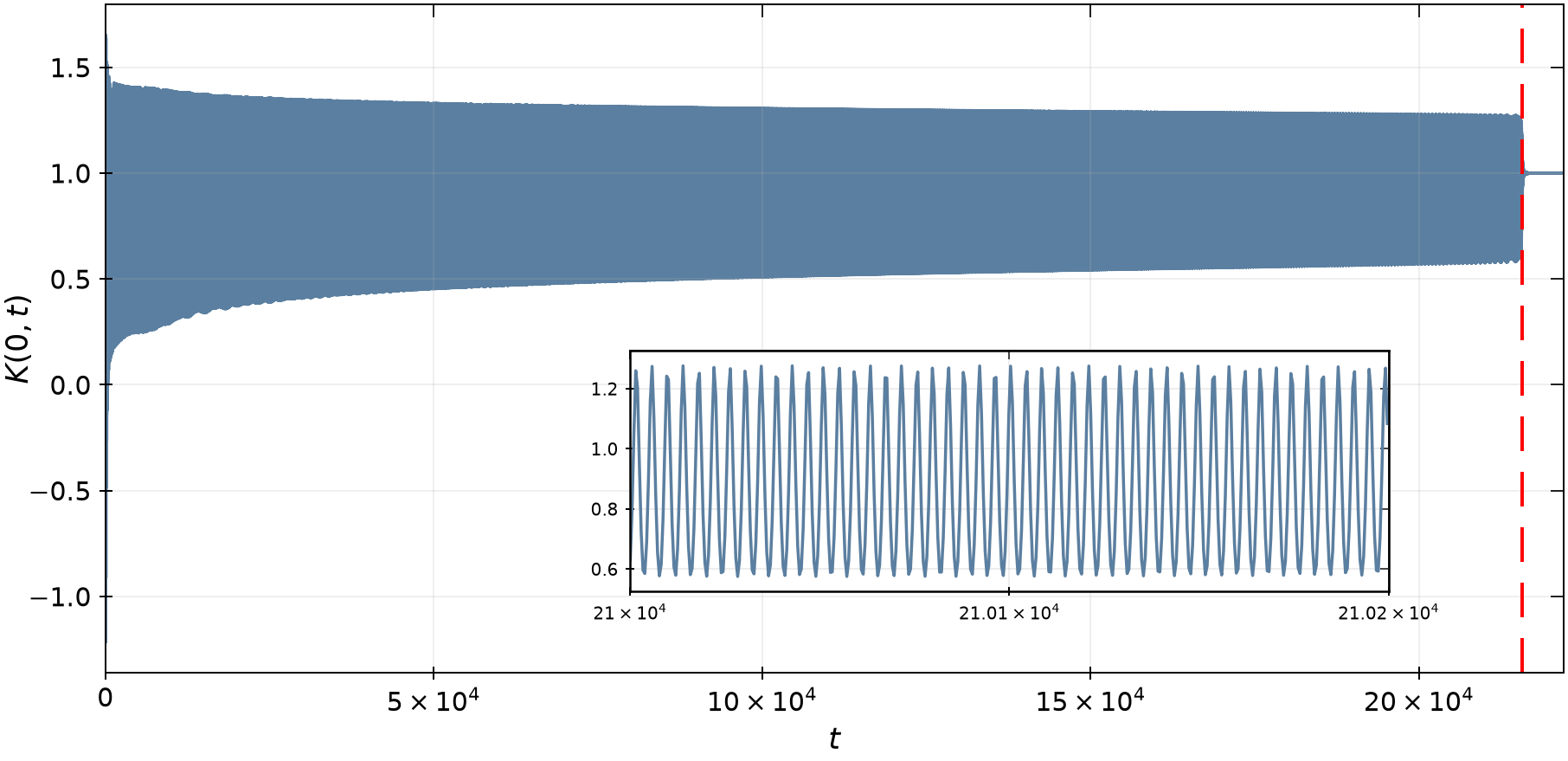} 
    \caption{
    Representative long-lived oscillon in the full five-field spherical ansatz. The plot shows the central Higgs component $K(0,t)$ for $\chi_6/m_H^2=0.4$, with initial data $A_K=0.2$, $R_0=2.4$, $A_H=-1$, and $R_H=14$.
The oscillon remains localized for $\tau\simeq 2.2\times 10^5\,m_W^{-1}$.}
    \label{full-oscillon}
    
    \vspace*{0.5cm}
    
    \includegraphics[width=\columnwidth]{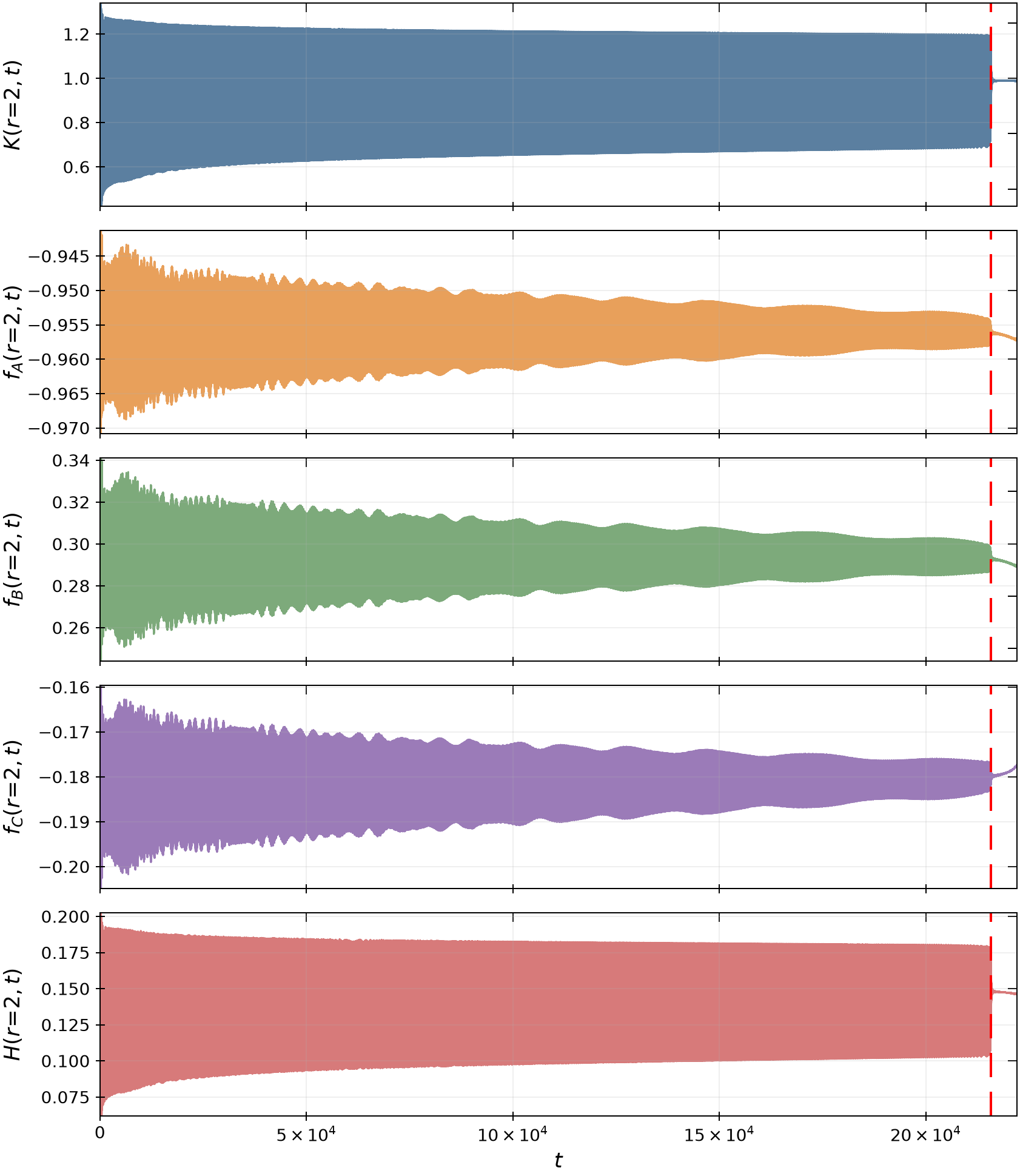} 
	\caption{
    Field evolution at fixed radius $r=2$ for the full five-field oscillon shown in Fig.~\ref{full-oscillon}. Away from the origin all five radial fields
$(f_A,f_B,f_C,H,K)$ exhibit nontrivial oscillatory dynamics. 
    } 
    \label{full-oscillon-10}
\end{figure}
\section{Oscillon in the full $SU(2)$ Higgs theory}

We now move beyond the consistent truncations discussed above and allow all five fields of the spherical ansatz, $(f_A,f_B,f_C,H,K)$, to evolve. We start from the Gaussian profile for $K$ in Eq.~\eqref{K-init} and add a localized perturbation in the other Higgs component,
\begin{equation}
H(r,0)=A_H\left ({r\over R_H}\right )e^{-r^2/R_H^2} \, .
\end{equation}
Through the coupled equations of motion, this perturbation excites the remaining gauge  components.

In the full five-field evolution, the Gauss constraint must also be maintained. Although we work in temporal gauge, $G=0$, the residual radial gauge structure of the ansatz is used to monitor and project the evolution back onto the constraint surface, as described in Appendix~\ref{app-gauge}.

In Fig. \ref{full-oscillon} we show a representative long-lived oscillon in the full five-field spherical ansatz, for $\chi_6/m_H^2=0.4$.
The figure shows the central value of the Higgs component, $K(r=0,t)$,
which remains localized and oscillatory for
$\tau\simeq 2.2\times 10^5\,m_W^{-1}$.
At the origin, regularity and
the equations of motion imply $H=f_B=0$ and $f_A=-1$, while away from
the origin all five fields display nontrivial oscillatory dynamics, as
shown in Fig.~\ref{full-oscillon-10}.
In particular,  the gauge
components lose amplitude on a shorter timescale than the Higgs components, while the long-lived configuration remains dominated by coherent Higgs-sector oscillation.
We emphasize that the long-lived oscillon is not a fine-tuned solution.
We find similar configurations for other values of $\chi_6$ and for
different choices of the initial profile; see Appendix~\ref{app-new-osc}.

\begin{figure}
         	\includegraphics[width=\columnwidth]{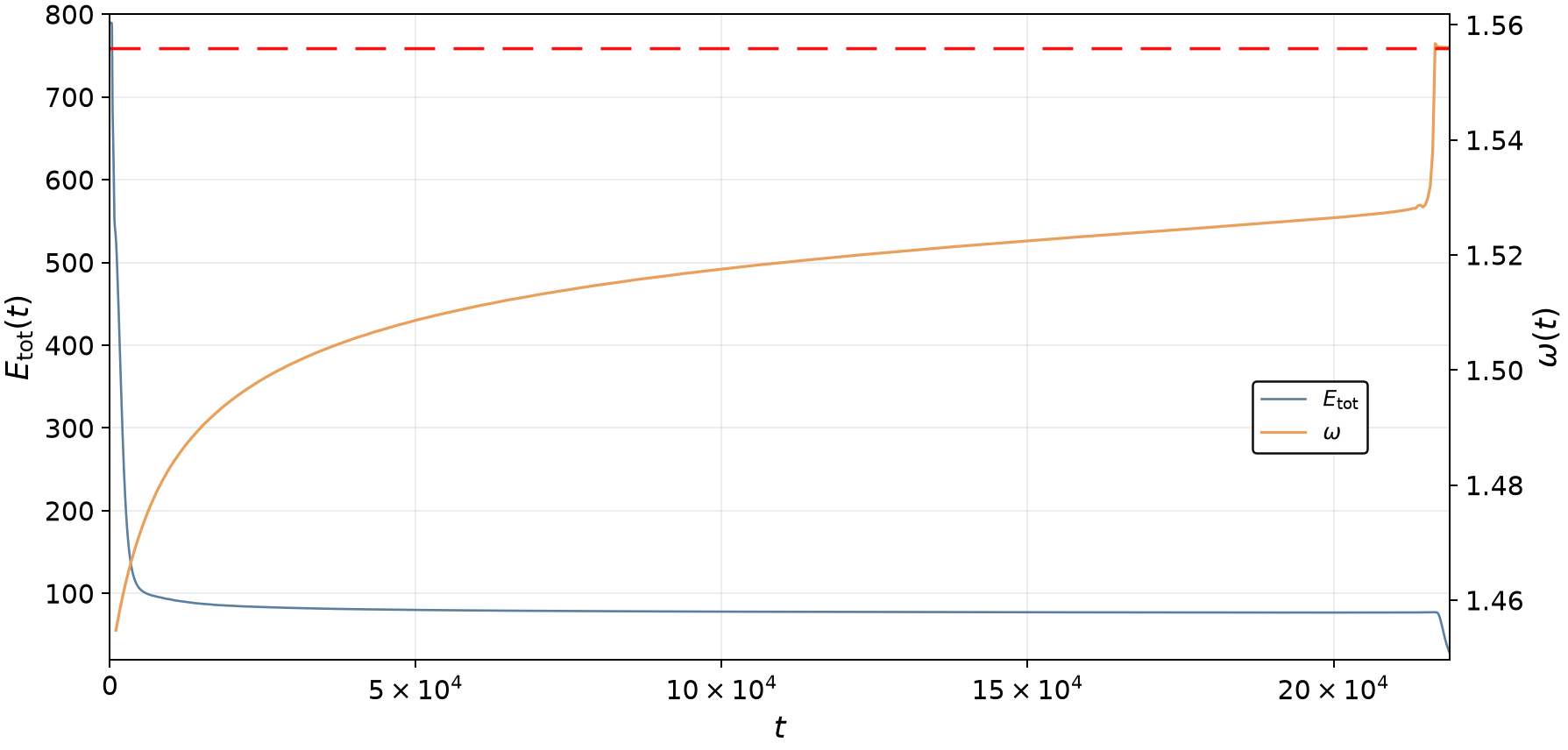} 
	\caption{Evolution of the energy and frequency of the SMEFT oscillon in Fig. \ref{full-oscillon}. Dashed line is the mass threshold in the Higgs sector $\omega=m_H$.} 
    \label{full-oscillon-freq}
\end{figure}

We note that the asymptotic values of the fields in Fig. \ref{full-oscillon-10} are $U(1)$-gauge equivalent to $f_A=-1, f_B=f_C=0$ and $K=1, H=0$, as discussed in Appendix~\ref{app-gauge}.

The fundamental frequency slowly drifts during the evolution, reflecting the continuous radiation of the oscillon; see Fig.~\ref{full-oscillon-freq}.
As in the single-field sector, the frequency of the Higgs components
$K$ and $H$ increases from $\omega\simeq 1.50$ at $t\simeq 2\times10^4$
to $\omega\simeq 1.556 = m_H$ shortly before decay. Thus the oscillon remains
below the Higgs radiation threshold, $\omega<m_H$, but above the gauge
threshold, $m_W=1$, as in Ref.~\cite{Farhi:2005rz}.

 Importantly, oscillon formation is robust. For the example in Fig.~\ref{full-oscillon-freq}, our chosen initial configuration has energy $E_{\rm init}=793$, while the oscillon energy is around $10\%$ of it, $E_{\rm osc}\simeq 88$--$79$. Thus, most of the initial energy is radiated away during the transient stage before the system settles into the localized oscillon. This provides further evidence that the SMEFT oscillon acts as an attractor of the classical evolution.

The mass (total energy) of the SMEFT oscillon is comparable but noticeably smaller than the mass of the SMEFT sphaleron. For the very long-lived oscillon shown above, we find $m_{osc}\approx 7.0-6.3$ TeV, while the corresponding sphaleron mass  is approximately $9$ TeV. This raises the possibility that the oscillon could appear as a long-lived intermediate state in sphaleron decay.

The SMEFT oscillon carries a negligible Chern--Simons charge at late times. During the initial transient, a small amount of charge is localized near the origin, $N_{\rm CS}\simeq 0.1$; see Appendix~\ref{app-CS}. Interestingly, this value is comparable to the Chern--Simons charge emitted as a spherical front in the decay of the
electroweak sphaleron~\cite{Hellmund:1991ub}.

\section{Conclusions}

We have investigated oscillons in the $SU(2)$ bosonic sector of the
Standard Model Effective Field Theory, focusing on the dimension-six
operator $\mathcal{O}_6=(\Phi^\dagger\Phi)^3$, one of 20 bosonic SMEFT operators. $\mathcal{O}_6$ modifies only the
Higgs potential and provides the simplest setting in which to test the
sensitivity of electroweak oscillons to SMEFT deformations.

Our main result is that the sextic interaction dramatically stabilizes
the oscillon. At the physical Higgs-to-$W$ mass ratio,
$m_H/m_W=1.556$, the inclusion of $\mathcal{O}_6$ increases the oscillon lifetime
by orders of magnitude and enlarges the basin of attraction for oscillon
formation. Long-lived configurations arise for values of the 
coefficient of $\mathcal{O}_6$ below the current experimental upper bound \cite{Elias-Miro:2013mua, Pomarol:2013zra,  Gan:2017mcv}, with the largest
enhancement occurring around $\chi_6/m_H^2\simeq 0.3$.

 In the strict single-field
truncation, the system reduces to a real scalar theory with a sextic
self-interaction. Opening a gauge-field direction does not immediately
destabilize the oscillon: The gauge perturbation relaxes toward its
vacuum value and the system settles back to a $K$-dominated
Higgs-channel oscillon. In the full five-field spherical ansatz, all
fields participate in the dynamics, but the long-lived oscillon persists.
This is nontrivial because the oscillon frequency lies below the Higgs
threshold but above the lighter gauge-field threshold.

The resulting SMEFT oscillon is not a fine-tuned solution. It forms
after substantial radiation from initial configurations whose energy is
well above the final oscillon energy, suggesting that it acts as an
attractor of the classical evolution. For the long-lived examples studied
here, the oscillon energy (mass) is $E_{\rm osc}\simeq 6.3$--$7.0\,{\rm TeV}$,
below the corresponding SMEFT sphaleron energy, $E_{\rm sph}\simeq
9\,{\rm TeV}$. This raises the possibility that the oscillon could appear
as a long-lived intermediate state in sphaleron decay.

Several questions remain open. The present analysis neglects the
$U(1)$ gauge field by setting $g'=0$; based on previous studies this is
not expected to qualitatively change the oscillon \cite{Farhi:2005rz, Graham:2006vy}, but it should be
checked explicitly. More broadly, many dimension-six SMEFT operators can
affect the bosonic solitonic sector \cite{Elias-Miro:2013mua, Gan:2017mcv}, including operators with derivative
interactions. Since the oscillon is sensitive to the Higgs-sector
deformation studied here, it may provide a useful probe of beyond-SM
effects encoded in SMEFT.

The SMEFT oscillon is a solution of the classical field equation and should be quantized. However, since it is a heavy, multi-particle state, its quantum state will be dominated by a classical solution with some semiclassical corrections. Importantly, it has been very recently shown that such corrections do not destabilize an oscillon \cite{Hertzberg:2010yz}, and it remains long-lived after quantization \cite{Evslin:2025hjt}.

 
\section*{Acknowledgements}
KS acknowledges financial support from the Polish National Science
Centre (Grant No. NCN 2021/43/D/ST2/01122). 
KS and AW are supported in part by the Spanish Ministerio de Ciencia e Innovacion (MCIN) with funding from the grant PID2023-148409NB-I00 MTM.
EIS is grateful for support from the  Jeff \& Gail Kodosky Endowed Chair at the University of Texas,
Austin. EIS also acknowledges
support from the U.S. Department of Energy, Office of Science, Office of High Energy Physics
program under Award Number DESC-0022021.

\newpage

\begin{center}
    \textbf{\large Supplement material}
\end{center}

\appendix
\section{Conventions}
We use the convention in which
\begin{align}
   W_\mu &=\frac{-i}{2}\tau^a W^a_\mu \, , \\
   W_{\mu \nu}&=\frac{-i}{2}\tau^a W^a_{\mu \nu} \, , \\
   D^\mu W_{\mu \nu}& =\frac{-i}{2}\tau^a (D^\mu W_{\mu \nu})^a \, ,
\end{align} 
where $\tau^a$ are Pauli matrices, obeying the usual relations $\operatorname{tr}(\tau^a\tau^b)=2\delta^{ab}$ and $[\tau^a,\tau^b]=2i\epsilon_{abc}\tau^c$. The gauge field is antihermitian $W^\dagger_\mu = -W_\mu$ as the components $W^a_{\mu \nu}$ are real. 

The action of the covariant derivative on the field tensor is defined as 
\begin{equation}
    D^\mu W_{\mu \nu} = \partial^\mu W_{\mu \nu} +g[W^\mu,W_{\mu \nu}] \, .
\end{equation}
Thus,
\begin{equation}
    W^a_{\mu \nu}=\partial_\mu W^a_\nu - \partial_\nu W^a_\mu +g\epsilon_{abc} W^b_\mu W^c_\nu
\end{equation}
and
\begin{equation}
   (D^\mu W_{\mu \nu})^a = \partial^\mu W^a_{\mu \nu}+g\epsilon_{abc} W^{b \mu} W^c_{\mu \nu} \, .
\end{equation}
Next, $\operatorname{tr} (W^{\mu \nu}W_{\mu \nu})=-(1/2) \left(W^{a\mu \nu}\right)^2$. 

\section{Equations of motion and energy}
\label{app-eom}

In the temporal gauge, $G=0$, the equations of motion for the radial fields $(f_A,f_B,f_C,H,K)$ in the SMEFT-deformed theory are as follows. Dots
and primes denote derivatives with respect to $t$ and $r$, respectively.
\begin{widetext}
\begin{align}
  \ddot f_A &= f_A'' - \frac{(f_A^2+f_B^2-1) f_A}{r^2} - m_W^2 \left[(H^2 + K^2) f_A + K^2 - H^2\right]
    - f_A f_C^2 + 2 f_B' f_C + f_B f_C', \\
  \ddot f_B &= f_B'' - \frac{(f_A^2+f_B^2-1) f_B}{r^2} - m_W^2 \left[(H^2 + K^2) f_B - 2HK\right]
    - f_B f_C^2 - 2 f_A' f_C - f_A f_C', \\
  \ddot f_C &=
    -\frac{2(f_A^2+f_B^2) f_C}{r^2} -m_W^2 (H^2 + K^2) f_C
    -2m_W^2(H'K-HK') -\frac{2(f_A' f_B-f_A f_B')}{r^2}, \\
  \ddot H &= H''+\frac{2H'}{r}
    -\frac{(f_A^2+f_B^2+1) H}{2r^2} +\frac{Hf_A+Kf_B}{r^2}
    -\biggl[
      \frac{m_H^2}{2}(H^2 + K^2 -1)
      + \frac{3\chi_6 m_W^2}{4}(H^2 + K^2 -1)^2
    \biggr] H \nonumber\\
  &\qquad
    +\frac{Kf_C}{r} -\frac{Hf_C^2}{4} +f_CK' +\frac{Kf_C'}{2}, \\
  \ddot K &= K''+\frac{2K'}{r}
    -\frac{(f_A^2+f_B^2+1) K}{2r^2} -\frac{Kf_A-Hf_B}{r^2}
    -\biggl[
      \frac{m_H^2}{2}(H^2 + K^2 -1)
      + \frac{3\chi_6 m_W^2}{4}(H^2 + K^2 -1)^2
    \biggr] K \nonumber\\
  &\qquad
    -\frac{Hf_C}{r} -\frac{Kf_C^2}{4} -f_CH' -\frac{Hf_C'}{2}.
\end{align}
\end{widetext}

Following~\cite{Matchev:2025ivr}, the total energy functional with $G$ kept explicit, is given by
\begin{equation}
    \label{eq:totalenergy}
    E \; = \; E_{\rm kin} + E_{\rm static} \, , 
\end{equation}
where the kinetic energy $ E_{\rm kin}$ and static energy $E_{\rm static}$ contributions are given by
\begin{widetext}
\begin{equation}
    E_{\rm kin} = \frac{4\pi}{g^2} \int_0^{\infty} \dd r \Biggl[ \dot{f}_A^2 \left( 1 + \frac{2f_B G}{\dot{f}_A}\right) + \dot{f}_B^2 \left( 1 - \frac{2f_A G}{\dot{f}_B}\right) + \frac{r^2}{2} \dot{f}_C^2 \left(1-2 \frac{G'}{\dot{f}_C} \right) 
    + 2 m_W^2 r^2 \left( \dot{H}^2 + \dot{K}^2 + G \left(\dot{H}K - H\dot{K}\right)\right) \Biggr] \, ,
 \label{eq:kinen}   
\end{equation}
\begin{equation}
\begin{aligned}
    & E_{\rm static} = \frac{4\pi}{g^2} \int_0^{\infty} \dd r \Biggl[ \left(f_A' + f_C f_B \right)^2 + \left(f_B' - f_C f_A \right)^2 + \frac{\left(f_A^2 + f_B^2 - 1 \right)^2}{2r^2} -\left(f_A^2 + f_B^2\right)G^2 - \frac{1}{2}r^2G'^2 \\
    & + 2 m_W^2 r^2 \bigg\{ \left(H' + \frac{1}{2}f_C K \right)^2 + \left(K' - \frac{1}{2} f_C H \right)^2  + \frac{1}{2r^2} \left(H f_A + K f_B - H \right)^2 + \frac{1}{2r^2} \left(K f_A - H f_B + K \right)^2 -\frac{1}{4} G^2(H^2 + K^2) \bigg\} \\
    & + \frac{(m_W m_H)^2}{2} r^2 \left(H^2 + K^2 -1 \right)^2  + \frac{\chi_6 m_W^4}{2} r^2 \left(H^2 + K^2 -1 \right)^3
    \Biggr] \, .
\label{eq:staten}   
\end{aligned}
\end{equation}

These expressions are completely general in regards to the gauge choice. In our numerics we worked in temporal gauge, fixing $G(r,t)=0$. Then, the formulas simplified to

\begin{equation}
    E_{\rm kin} = \frac{4\pi}{g^2} \int_0^{\infty} \dd r \Biggl[ \dot{f}_A^2  + \dot{f}_B^2  + \frac{r^2}{2} \dot{f}_C^2
    + 2 m_W^2 r^2 \left( \dot{H}^2 + \dot{K}^2\right) \Biggr] \, ,
 \label{eq:kinen}   
\end{equation}
\begin{equation}
\begin{aligned}
    & E_{\rm static} = \frac{4\pi}{g^2} \int_0^{\infty} \dd r \Biggl[ \left(f_A' + f_C f_B \right)^2 + \left(f_B' - f_C f_A \right)^2 + \frac{\left(f_A^2 + f_B^2 - 1 \right)^2}{2r^2}  \\
    & + 2 m_W^2 r^2 \bigg\{ \left(H' + \frac{1}{2}f_C K \right)^2 + \left(K' - \frac{1}{2} f_C H \right)^2  + \frac{1}{2r^2} \left(H f_A + K f_B - H \right)^2 + \frac{1}{2r^2} \left(K f_A - H f_B + K \right)^2  \bigg\} \\
    & + \frac{(m_W m_H)^2}{2} r^2 \left(H^2 + K^2 -1 \right)^2  + \frac{\chi_6 m_W^4}{2} r^2 \left(H^2 + K^2 -1 \right)^3
    \Biggr] \, .
\label{eq:staten-G0}   
\end{aligned}
\end{equation}
\end{widetext}

\section{Residual $U(1)$ gauge symmetry}
\label{app-gauge}
\begin{figure}
	\includegraphics[width=\columnwidth]{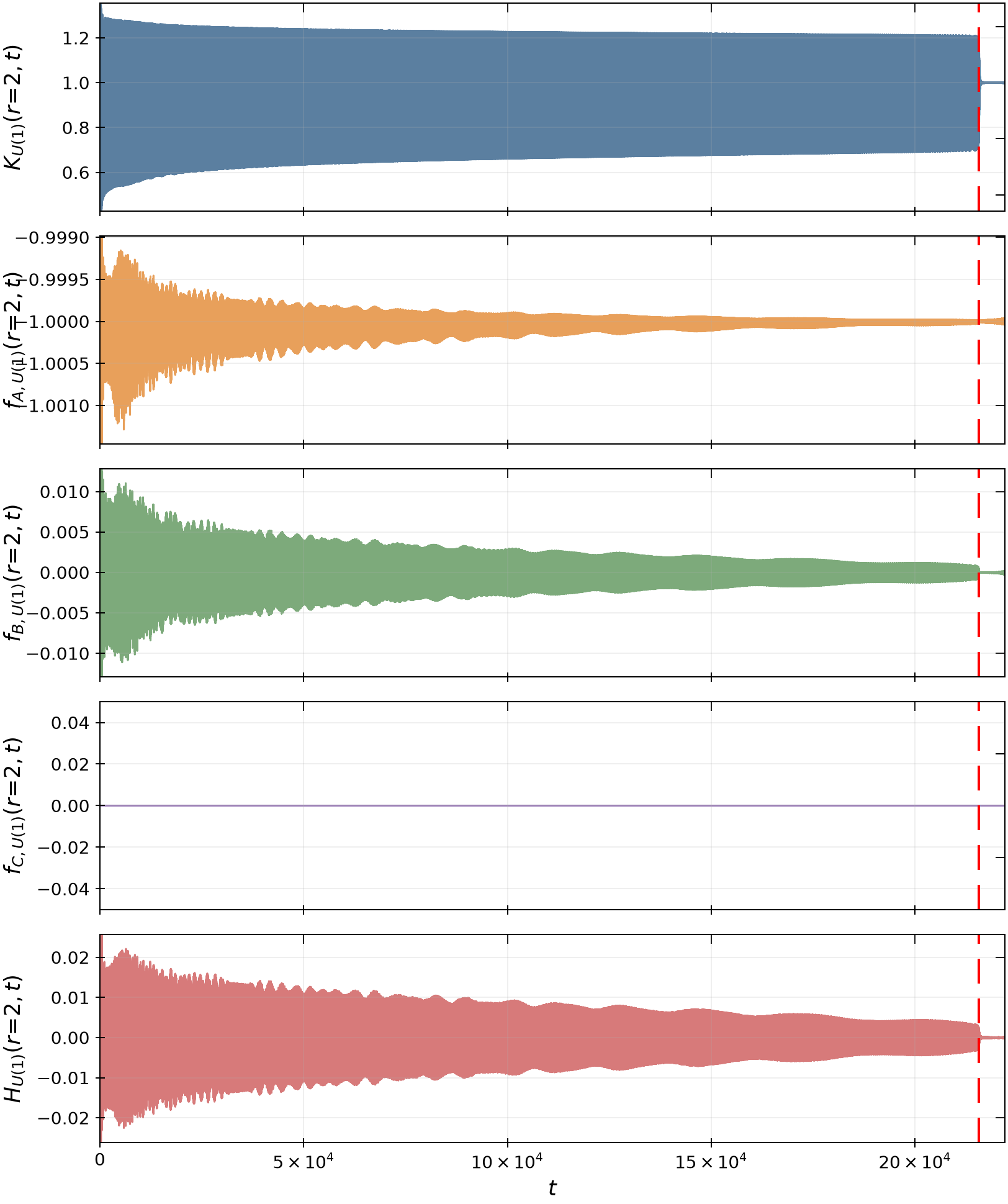} 
    \caption{The fields at $r=2$ of the five-channel oscillon with a long lifetime from Fig. \ref{full-oscillon-10}. Here the functions are gauge-shifted to have $f_C\equiv 0$.}
    \label{full-oscillon-new-gauge}
\end{figure}


The radial ansatz does not fix the gauge completely. 
It preserves a
residual radial $U(1)$ gauge structure parametrized by a function $\theta(r,t)$, under which the fields transform as
\begin{align}
\begin{pmatrix}
G \\
f_C
\end{pmatrix}
&\rightarrow
\begin{pmatrix}
G \\
f_C
\end{pmatrix}
+
\begin{pmatrix}
\dot{\theta}(r,t) \\
\theta'(r,t)
\end{pmatrix} \, ,  \\
\begin{pmatrix}
f_A \\
f_B
\end{pmatrix}
&\rightarrow
\begin{pmatrix}
\cos\theta(r,t) & -\sin\theta(r,t) \\
\sin\theta(r,t) & \cos\theta(r,t)
\end{pmatrix}
\begin{pmatrix}
f_A \\
f_B
\end{pmatrix} \, , \\
\begin{pmatrix}
H \\
K
\end{pmatrix}
&\rightarrow
\begin{pmatrix}
\cos\left(\frac{\theta(r,t)}{2}\right) & -\sin\left(\frac{\theta(r,t)}{2}\right) \\
\sin\left(\frac{\theta(r,t)}{2}\right) & \cos\left(\frac{\theta(r,t)}{2}\right)
\end{pmatrix}
\begin{pmatrix}
H \\
K
\end{pmatrix} \, . 
\end{align}
The prime denotes a radial derivative, $\theta'(r,t) \equiv \partial \theta(r,t)/\partial r$, and the dot denotes a time derivative, $\dot{\theta}(r,t) \equiv \partial \theta(r,t)/\partial t$. The function $\theta(r,t)$ parameterizes local $\mathrm{SU}(2)$ gauge transformations that preserve the structure of the ansatz. 

After imposing temporal gauge, $G=0$, only transformations with $\dot\theta=0$ preserve the gauge condition. Nevertheless, the associated
Gauss constraint must be monitored during the full five-field evolution.
In terms of the ansatz variables it reads

\begin{widetext}
    \begin{equation}
        G''+ \frac{2G'}{r}  - \dot{f}'_C-\frac{2\dot{f}_C}{r} -\frac{2}{r^2} \left( f_A^2+f_B^2\right)G +\frac{2}{r^2} \left( f_A \dot{f}_B-f_B\dot{f}_A\right) + 2m_W^2 \left(H\dot{K}-K\dot{H} \right) - m^2_W \left( H^2+K^2 
        \right)G=0 \, .
    \end{equation}
\end{widetext}

In the numerical evolution, small violations of this constraint can be generated by discretization errors. We correct them by a projection step: the radial elliptic constraint is solved and the fields are projected back onto the constraint surface while maintaining the temporal-gauge condition $G=0$.


The same residual gauge structure is useful in post-processing. At each time slice one may choose $\theta'(r,t)=-f_C(r,t)$, so that the displayed fields satisfy $f_C\equiv0$. This representation makes the approach to the standard vacuum manifest: $K$ and $H$ oscillate around $1$ and $0$, while $f_A$ and $f_B$ oscillate around $-1$ and $0$, respectively; see
Fig.~\ref{full-oscillon-new-gauge}.


\newpage
\section{SMEFT oscillon from alternative initial data}
\label{app-new-osc}
\begin{figure}
	\includegraphics[width=\columnwidth]{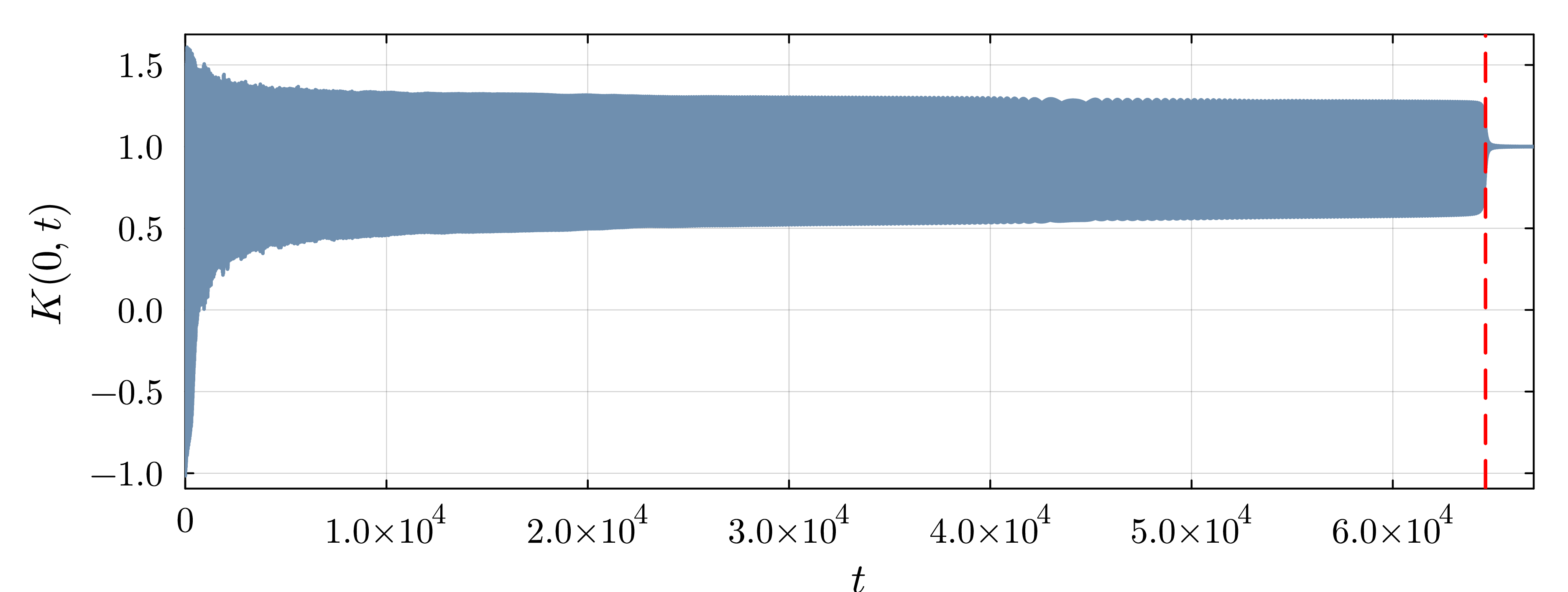} 
    \caption{
    Long-lived full five-field SMEFT oscillon obtained from an alternative
choice of initial data. The plot shows $K(0,t)$ for $\chi_6/m_H^2=0.39$, with $A_K=-2.0$, $R_0=2.75$, $A_A=0.1$, $R_A=2.3$, $A_H=10^{-5}$, and $R_H=2.85$.
The lifetime is $\tau\simeq 6.3\times10^4\,m_W^{-1}$.}
    \label{full-oscillon-new}
    
    \includegraphics[width=\columnwidth]{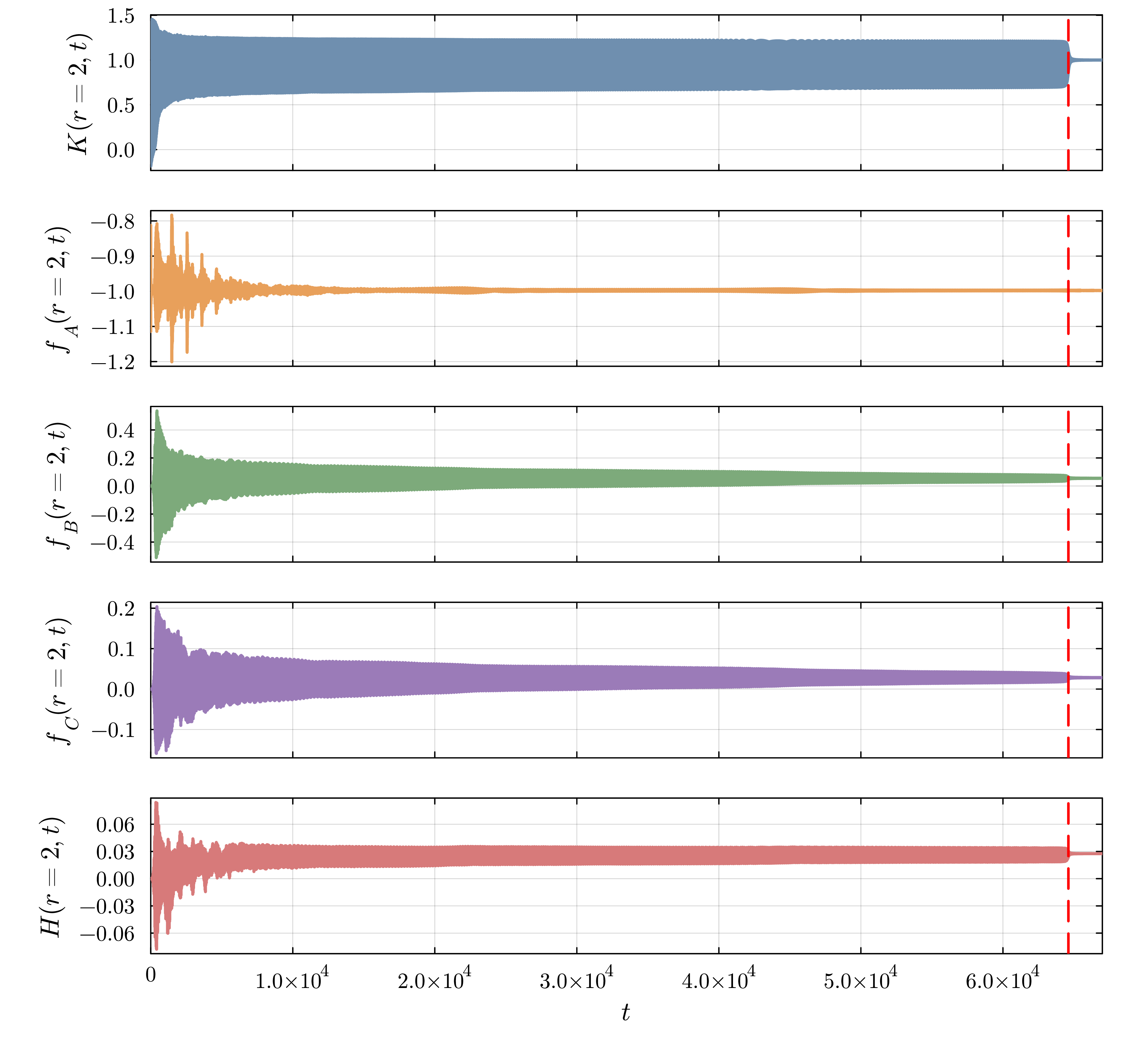} 
	\caption{
 Field evolution at fixed radius $r=2$ for the alternative full five-field oscillon shown in Fig.~\ref{full-oscillon-new}.}
    \label{full-oscillon-10-new}
\end{figure}

\begin{figure}
         	\includegraphics[width=\columnwidth]{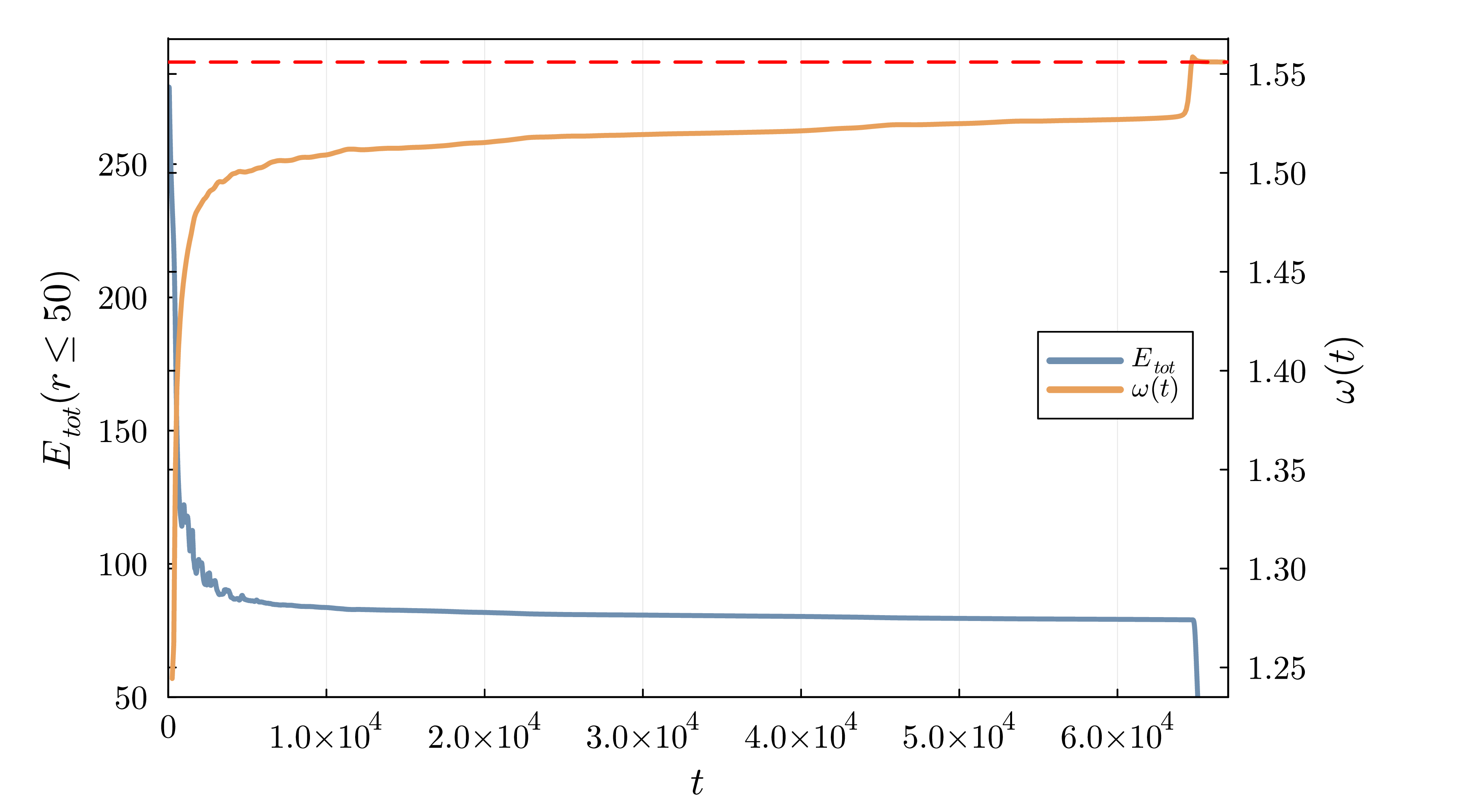} 
	\caption{
Localized energy and frequency evolution for the alternative full five-field oscillon shown in Fig.~\ref{full-oscillon-new}. The energy is measured inside the region with $r<50$. The dashed line denotes the Higgs radiation threshold,
$\omega=m_H$. The energy and frequency are comparable to those of the
longer-lived configuration shown in Fig.~\ref{full-oscillon-freq}.}
    \label{full-oscillon-new-freq}
\end{figure}

To illustrate that the full five-field SMEFT oscillon is not a fine-tuned solution, tied to a
single choice of initial profile, we consider an alternative perturbation
of the $H$ field. We take the initial pulses in the $K$ and $f_A$
components as in Eqs.~\eqref{K-init} and~\eqref{fA-init}, and supplement them with a different perturbation in the $H$ field
\begin{equation}
    H(r,0)=A_H r^2 e^{-\frac{r^2}{R_H^2}}.
\end{equation}
For $\chi_6 = 0.39 m^2_H$ and $A_K=-2.0,
R_0 = 2.75, A_A = 0.1, R_A = 2.3, A_H = 10^{-5}, R_H = 2.85$, we find a long-lived oscillon with $\tau \approx 63 \; 000$; see Fig. \ref{full-oscillon-new} and Fig. \ref{full-oscillon-10-new}.

The  energy of this oscillon and its frequency evolution are comparable to those of the long-lived example discussed in the main text; see Fig.~\ref{full-oscillon-new-freq}. We
find similar configurations for other  choices of the initial
parameters.



\begin{figure*}
    \includegraphics[width=\columnwidth]{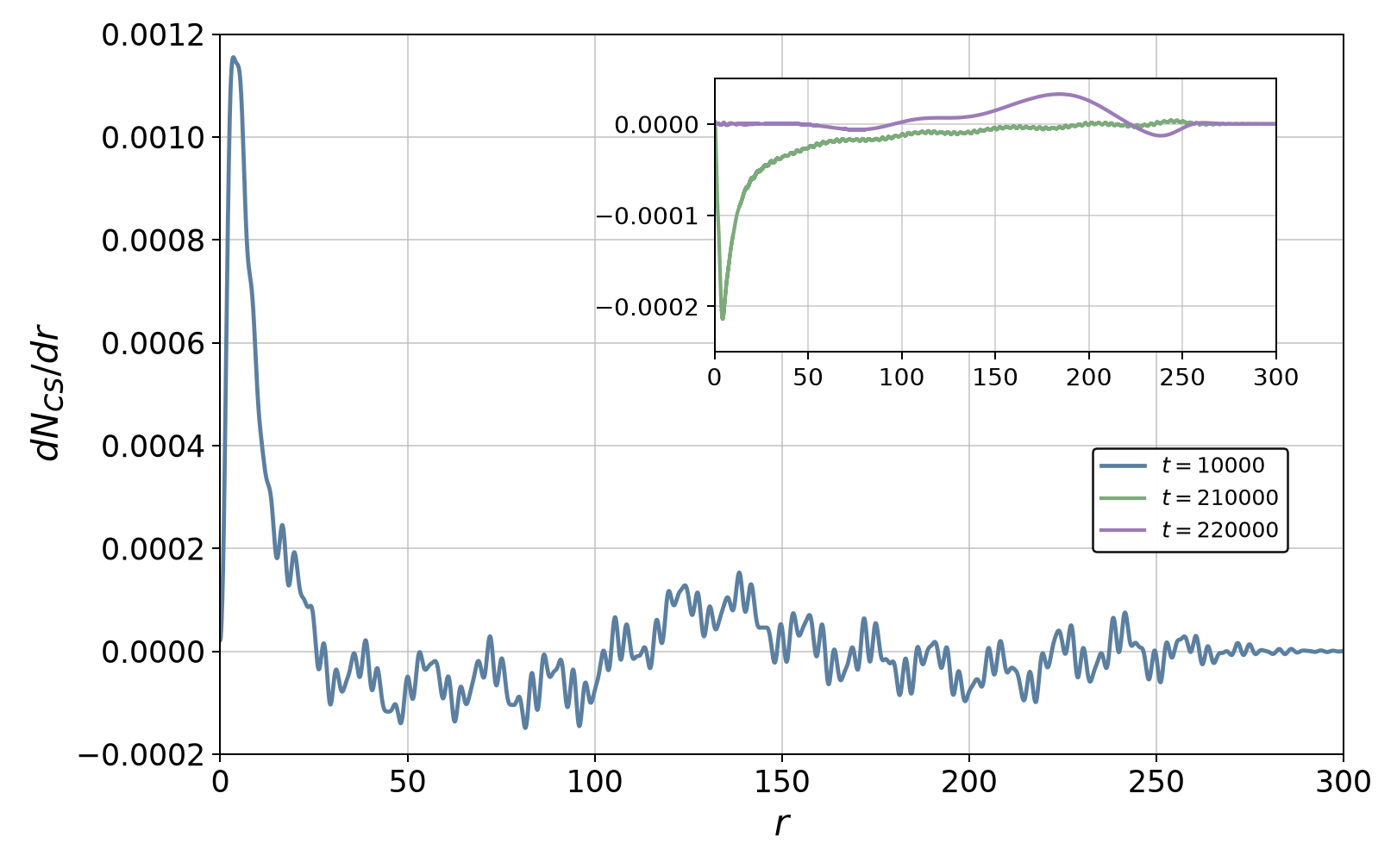}
    \includegraphics[width=\columnwidth]{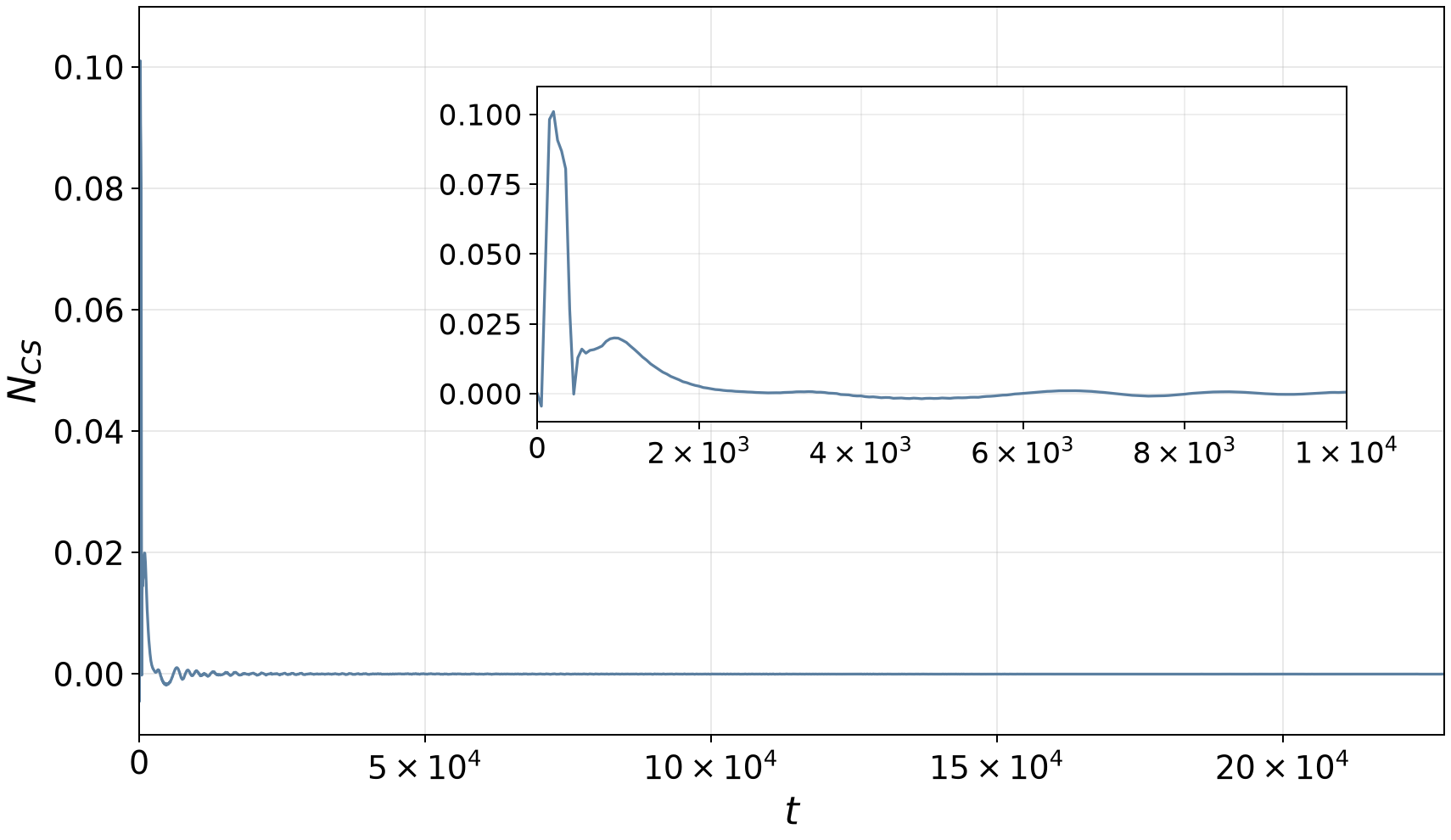}
    \caption{Left: evolution of the density of the Chern--Simons charge of the SMEFT oscillon from Fig. \ref{full-oscillon}. Right: evolution of the Chern--Simons charge localized on the SMEFT oscillon.} 
    \label{full-oscillon-CS}
\end{figure*}

\newpage
\section{Chern--Simons charge}
\label{app-CS}

The Chern--Simons current reads

\begin{equation}
    \mathcal{K}^\mu=\frac{g^2}{16\pi^2} \epsilon^{\mu \nu \rho \sigma} \left( 
   W^a_\nu \partial_\rho W^a_\sigma - \frac{g}{3} \epsilon^{abc} W^a_\nu W^b_\rho W^c_\sigma \right).
\end{equation}
It gives rise to a topological charge
\begin{equation}
    N_{CS}=\frac{g^2}{32\pi^2} \int d^4 x \mbox{Tr} \left( W_{\mu \nu} ^*W^{\mu \nu}\right),
\end{equation}
where $^*W^{a \mu \nu}=(1/2)\epsilon^{\mu \nu \rho \sigma} W^a_{\rho \sigma}$ is the dual field tensor. It can be written as
\begin{equation}
  \left.  N_{CS}=\int d^3x\mathcal{K}^0 \right|_{t=t_0}^{t=-\infty}+\int_{-\infty}^{t_0} \int_{S_\infty}\vec{\mathcal{K}}\cdot d\vec{S},
\end{equation}
where $S_\infty$ is the surface at infinity. The surface integral vanishes for sufficiently rapid decaying solutions. Then, assuming that $\mathcal{K}^0(t=-\infty)=0$, we find 
\begin{equation}
    N_{CS}(t_0)=\int d^3x\mathcal{K}^0(t_0)
\end{equation} In the radial ansatz it takes the following form
\begin{widetext}
\begin{equation}\label{n}
N_{CS} = \frac{1}{2\pi} \left(f_B(r_{max}) - f_B(r_{min}) + \int_{r=r_{min}}^{r_{max}} (f_A' f_B - f_A f_B') \dd r \right).
\end{equation}
\end{widetext}
In principle, there is also a term 
\begin{equation}
   N_q= \frac{q-\sin(q)}{2\pi},
\end{equation}
where $q$ is the parameter that labels the pure gauge vacuum
\begin{equation}
    U=\mbox{exp}\left(\frac{iq}{2} \frac{\vec{\tau} \cdot\vec{r}}{r} \right).
\end{equation}
For the oscillon $q=0$.

In the case of the long-lived SMEFT oscillon from Fig. \ref{full-oscillon}, initially, for $t<500$, there is a small but very well visible amount of the CS charge localized in the origin, $N_{CS}\approx 0.1$; see Fig. \ref{full-oscillon-CS}. This localized charge is radiated out in the form of charge waves.

\section{Numerical implementation}
\label{app-numerics}



The single-field and gauge-perturbed truncations of Section~\ref{1-field} were evolved on a uniform radial grid using a fixed-step fourth-order Runge-Kutta scheme, with regularity conditions at the origin and outgoing boundary conditions at large $r$. The oscillon lifetime was defined from the localized energy: after the initial transient, a plateau value was determined by a moving average, and $\tau$ was taken to be the first time at which the energy inside the oscillon region dropped by $90\%$ relative to this plateau. The time-dependent frequency was extracted using a sliding-window FFT of
$K(0,t)$.
\\
The full spherically symmetric $SU(2)$--Higgs system was evolved in the five-field sector $(f_A,f_B,f_C,H,K)$ on a uniform radial grid. The second-order equations were integrated with a fixed-step leapfrog/velocity-Verlet scheme and finite-difference spatial derivatives. A sponge layer near the outer boundary absorbed outgoing radiation. The Gauss constraint was monitored throughout the evolution and corrected, when necessary, by projecting back onto the constraint surface through the associated radial elliptic equation.

\newpage

\bibliography{ewsphst}

\end{document}